\documentclass[twocolumn,trackchanges]{aastex62}

\usepackage{amsmath}
\usepackage{natbib}
\usepackage{url}
\usepackage{color}

\newcommand{\pkg}[1]{\texttt{#1}} 
\newcommand{\kword}[1]{\texttt{#1}} 
\newcommand{\sol}{$_{\odot}$}

\accepted{March 23, 2021}
\submitjournal{ApJ}

\shorttitle{The NHS Catalog: Quiescent Galaxies in SHELA}
\shortauthors{Stevans et al.}

\begin{document}

\turnoffedit1
\turnoffedit2 
\turnoffedit3

\title{The NEWFIRM HETDEX Survey: Photometric Catalog and a Conservative Sample of Massive Quiescent Galaxies at $z=3-5$ over 17.5 deg$^2$ in the SHELA Field}


 \correspondingauthor{Matthew L. Stevans}
 \email{stevans@utexas.edu}

 \author[0000-0001-8379-7606]{Matthew L. Stevans}
 \affiliation{Department of Astronomy, University of Texas at Austin, Austin, TX 78712}
 
  \author[0000-0001-8519-1130]{Steven L. Finkelstein}
 \affiliation{Department of Astronomy, University of Texas at Austin, Austin, TX 78712}

\author[0000-0003-4032-2445]{Lalitwadee Kawinwanichakij}
\affiliation{Department of Physics and Astronomy, Texas A\&M University, College
Station, TX, 77843-4242 USA}
\affiliation{George P.\ and Cynthia Woods Mitchell Institute for
  Fundamental Physics and Astronomy, Texas A\&M University, College
  Station, TX, 77843-4242 USA}
\affiliation{LSSTC Data Science Fellow}  

\author{Isak Wold} 
 \affiliation{Astrophysics Science Division, Goddard Space Flight Center, 8800 Greenbelt Road, Greenbelt, Maryland, 20771}

\author[0000-0001-7503-8482]{Casey Papovich}
\affiliation{Department of Physics and Astronomy, Texas A\&M University, College
Station, TX, 77843-4242 USA}
\affiliation{George P.\ and Cynthia Woods Mitchell Institute for
  Fundamental Physics and Astronomy, Texas A\&M University, College
  Station, TX, 77843-4242 USA}

\author{Rachel S. Somerville}
 \affiliation{Center for Computational Astrophysics, Flatiron Institute, 162 5th Avenue, New York, NY, 10010, USA}
 \affiliation{Department of Physics and Astronomy, Rutgers, The State University of New Jersey, Piscataway, NJ, 08854, USA}
 
\author{L.\ Y.\ Aaron Yung}
 \affiliation{Department of Physics and Astronomy, Rutgers, The State University of New Jersey, Piscataway, NJ, 08854, USA}
 \affiliation{Center for Computational Astrophysics, Flatiron Institute, 162 5th Avenue, New York, NY, 10010, USA}
 
\author{Sydney Sherman}
 \affiliation{Department of Astronomy, University of Texas at Austin, Austin, TX 78705}
 
\author{Robin Ciardullo}
 \affiliation{Department of Astronomy \& Astrophysics, The Pennsylvania State University, University Park, PA 16802}
 \affiliation{Institute for Gravitation and the Cosmos, The Pennsylvania State University, University Park, PA 16802}

\author{Romeel Dav\'{e}}
\affiliation{Institute for Astronomy, Royal Observatory, University of Edinburgh, Edinburgh EH9 3HJ, UK}

\author{Jonathan Florez}
 \affiliation{Department of Astronomy, University of Texas at Austin, Austin, TX 78705}
 
\author{Caryl Gronwall}
 \affiliation{Department of Astronomy \& Astrophysics, The Pennsylvania State University, University Park, PA 16802}
 \affiliation{Institute for Gravitation and the Cosmos, The Pennsylvania State University, University Park, PA 16802}

\author{Shardha Jogee}
 \affiliation{Department of Astronomy, University of Texas at Austin, Austin, TX 78705}
 
\begin{abstract}
We present the results of a deep $K_s$-band ($2.1 \mu$m) imaging survey of the {\it Spitzer}/HETDEX Exploratory Large Area (SHELA) field using the NEWFIRM  near-infrared (NIR) camera on the KPNO Mayall 4-m telescope. This NEWFIRM HETDEX Survey (NHS) reaches a $5\,\sigma$ depth of 22.4 AB mag ($2\arcsec$-diameter apertures corrected to total), is $\sim$50\% and 90\% complete at $K\sim$22.65 and $K\sim$22.15, respectively, and covers 22~deg$^2$ of the 24~deg$^2$ SHELA {\it Spitzer}/IRAC footprint (within ``Stripe 82'' of the Sloan Digital Sky Survey).  We present a $K_s$-band-selected catalog which includes deep $ugriz$ imaging from the Dark Energy Camera and 3.6 and 4.5~$\mu$m imaging from {\sl Spitzer\/}/IRAC, with forced-photometry of 1.7 million sources across 17.5 deg$^2$.   The large area and moderate depth of this catalog enables the study of the most massive galaxies at high redshift, and minimizes uncertainties associated with counting statistics and cosmic variance. As a demonstration, we derive stellar masses (M$_{\ast}$) and star-formation rates (SFRs) for candidate galaxies at $3 \lesssim z \lesssim 5$, and select a conservative sample of nine candidate massive (M$_{\ast}$ $> 10^{11}$~M$_{\odot}$) quiescent galaxies, which have measured SFRs significantly below the main-sequence at this redshift.  Five are ultra-massive with M$_{\ast}$ $> 10^{12}$, though uncertainties in IRAC blending, gravitational lensing, or AGN emission could result in true masses which are lower. Simulations predict that these galaxies should be extremely rare, thus we conclude by discussing what physical processes in models could be altered to allow the formation of such massive quiescent galaxies at such early times.
\end{abstract}



\section{Introduction} \label{sec:intro}

It is well-established observationally that massive galaxies in the local Universe have a strongly bimodal color distribution \citep{baldry04,bell04}. This is interpreted as indicating the presence of a star forming population of galaxies, which generally populate a relatively tight relation between stellar mass and star formation rate (SFR), and a population of ``quiescent'' galaxies that contain predominantly old stellar populations \citep{brinchmann04}. It is also well-established that the stellar mass and number of galaxies contained in this quiescent population has increased significantly over cosmic time since $z\sim 2$ \citep[e.g.,][]{kriek06,muzzin13,stefanon13,brennan15}, while the mass in the star forming population has remained nearly constant, indicating that some process (which has come to be commonly referred to as ``quenching'') is causing star forming galaxies to stop forming new stars and become ``red and dead''. The observed predominance of quiescent massive galaxies has historically been a challenge for theoretical models of galaxy formation, as the massive halos that host these objects are expected to experience rapid cooling flows of fresh gas to the center, where it should fuel star formation. Numerous physical mechanisms that could cause or contribute to quenching have been considered, but there is now a consensus that feedback processes related to radiation and kinetic energy injected by accreting supermassive black holes (AGN feedback) are likely the dominant mechanisms leading to long-lived quenching  \citep[see discussion in][and references therein]{somerville15a}. Thus, constraining when the first quenched galaxies appeared in the Universe, and how common they are at early epochs, can provide critical constraints on when and where the first supermassive black holes formed, and how they impact their surroundings, which are key open questions in galaxy formation. Most recent simulations of galaxy formation are able to reproduce observed fractions of massive quenched galaxies in the local Universe \citep[e.g.,][]{kimm09,brennan15,nelson18,hahn19}, but reproducing as many massive quenched galaxies as observed at higher redshifts $z\gtrsim 2$ appears to pose a challenge for these models \citep{brennan15,merlin19}.  

Such an investigation requires an unbiased census of massive galaxies over cosmic time. A number of observational studies have measured the number density of quiescent galaxies or the quiescent galaxy fraction out to high redshift \citep[e.g.,][]{dunlop07, marchesini10, muzzin13, straatman14, spitler14, brennan15,stefanon15, davidzon17, merlin19}, but the results differ substantially from one study to another. This may be in part because there is no clear universal definition of how to identify ``true'' quiescent galaxies at different epochs, but is likely also because these objects are quite rare, and these surveys have typically probed relatively small volumes, making counting statistics and field-to-field variance a substantial source of uncertainty. Additionally, there has been progress spectroscopically confirming massive quiescent galaxy candidates at $z>3$ \citep[e.g.,][]{glazebrook17,schreiber18,tanaka19,forrest20a,valentino20} and in ruling out obscured star formation with deep deep Atacama Large Millimeter/submillimeter Array (ALMA) observations \citep{santini19}, but more work is needed in both of these areas.

In this paper, we perform a systematic search for massive quiescent galaxies in the early universe ($3<z<5$) using deep multi-wavelength imaging across a wide area within the Spitzer/HETDEX Exploratory Large-Area (SHELA) field \citep{papovich16}. SHELA is a legacy field covering an area of $\sim$ 24 deg$^2$ within the footprints of the SDSS ``Stripe 82'' and the Hobby-Eberly Telescope Dark Energy Experiment \citep[HETDEX;][]{hill16}. HETDEX is measuring the effects of dark energy on the expansion rate of the universe via the baryonic acoustic oscillation (BAO) scale length by measuring spectroscopic redshifts of one million Lyman-$\alpha$ emitting galaxies at $z=1.9-3.5$. The SHELA dataset includes deep (22.6 AB mag, 50\% completeness) 3.6 $\mu$m and 4.5 $\mu$m imaging from \textit{Spitzer}/IRAC \citep{papovich16}, $u^{\prime}g^{\prime}r^{\prime}i^{\prime}z^{\prime}$ imaging from the Dark Energy Camera over 18 deg$^2$ (DECam; \citealt{wold19}), VISTA $J$ and $K_{s}$ photometry from the VICS82 survey \citep{geach17}, and a growing database of full-field IFU spectroscopy from HETDEX (presently $\sim$20\% complete).


Such a survey depends sensitively on the depth of its NIR imaging, and thus here we present the NEWFIRM HETDEX Survey\footnote{NHS high-level image products will be available here: \url{https://www.noao.edu/survey-archives/}} (NHS), a moderately deep $K_{s}$ (2.1 $\mu$m) near-infrared (NIR) imaging survey with NEWFIRM on the KPNO Mayall 4m Telescope. The NHS adds to SHELA $K_{s}$-band imaging (5$\sigma$ $K_{s}=$22.4 mag AB in a 2$"$-diameter aperture) across 22 deg$^2$, which is important for reducing the fraction of catastrophic errors in photometric redshifts and measuring robust star-formation rates (SFRs) by breaking the age-dust degeneracy. We present the NHS imaging and photometric catalog, along with a catalog of galaxy properties from SED-fitting for massive galaxies (M/M$_{\odot}$ $> 10^{11}$) at $z=3-5$. \edit2{We highlight the utility of this catalog for exploring rare objects by identifying a conservative sample of nine candidate massive galaxies (M/M$_{\odot}$ $> 10^{11}$) which are quiescent (forming stars at a rate well below the main-sequence at that stellar mass).}

This paper is organized as follows. The NHS observations and survey
strategy are described in Section \ref{sec:observations}.  The NHS
data reduction, calibration, weighted combination, and
position-dependent modeling of the point spread function (PSF) are
described in Section \ref{sec:reduction}. NHS source cataloging,
including aperture photometry and error estimation, are described in
Section \ref{sec:cataloging}. We describe the matching to the
ancillary datasets (i.e., IRAC, DECam, and VICS82) in Section
\ref{sec:matching}. Our method for deriving galaxy properties with
\pkg{eazy-py} is described in Section \ref{sec:eazypy}. Our sample
selection, contamination estimation, and measurement of the quiescent
fraction are presented in Section \ref{sec:results}. The implications
of our results are discussed in Section \ref{sec:discussion}. We
summarize our work and discuss future work in Section
\ref{sec:summary}. Throughout this paper we assume a Planck 2014
cosmology, with H$_0 = 67.8$ km s$^{-1}$ Mpc$^{-1}$, $\Omega_M =
0.307$ and $\Omega_{\Lambda} = 0.693$ \citep{planck13}. All magnitudes
given are in the AB system \citep{oke83}.

\section{Observations and Survey Strategy} \label{sec:observations}
We observed the SHELA Field with the Mayall 4m Telescope at Kitt 
Peak National Observatory using the NOAO Extremely Wide-Field Infrared 
Imager \citep[NEWFIRM;][]{probst04} instrument and $K_{s}$ broadband filter. The NEWFIRM detector is comprised of four, 2048 x 2048 ORION InSb arrays 
with localized responsivity defects and a pixel scale of  0.4$''$/pixel. 
The field of view is 27.6$'$ x 27.6$'$ including a 35$''$ wide cross-shaped 
gap between the four detectors. Images were 
taken over 93 nights in semesters 2014B, 2015B, 2016B, and 2017B 
under NOAO Survey Program \#13B-0236 (PI: Finkelstein).

\begin{figure*}[!tbp]
\centering
\includegraphics[scale=0.35,angle=0]{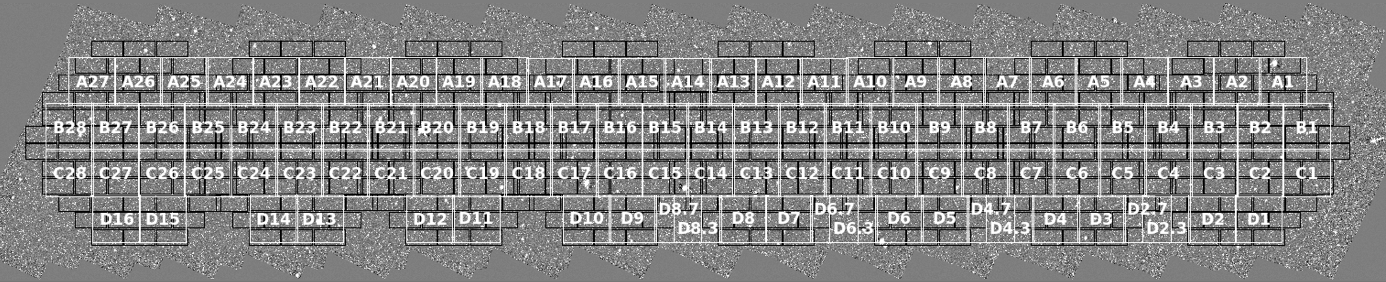}
\caption{Layout of the NHS SHELA field NEWFIRM pointings (white squares; 27.6$'$ x 27.6$'$ field of view) overplotted on the IRAC data (grey) and the DECam pointings (black; individual DECam CCDs are traced). The left-most two (of eight) DECam pointings were incomplete at the time of this analysis, thus are not used here. The NHS layout was designed to maximize the area of overlap with DECam data. Ten additional overlapping NEWFIRM pointings were observed in row D after the nominal exposure time (160 minutes) was reached in the initially planned 99 pointings.}
\label{Fig:field}
\end{figure*}

The field was observed in a pattern of 109 tiled pointings in four rows that maximizes
the area of overlap with the existing SHELA DECam observations \citep{wold19} as shown in Figure \ref{Fig:field}. Each tile is labeled with a letter (A-D) and a number (0-28). 
The letter indicates the row (row A having the highest declination and D the lowest),
while the number indicates the relative position in right ascension starting at 1 for the lowest right ascension. The tiles overlap each other by about 1 arcmin. After the nominal exposure time was reached in the initially planned
99 pointings, 10 supplemental pointings were added to row D to fill in the gaps
between the large circular DECam pointings where data do not overlap.

To reach the target depth of $K_{s}=22.7$ mag across the field \edit2{(in 2$^{\prime\prime}$ apertures)}, an exposure 
time of 160 minutes per pointing was planned. Each exposure was
60 seconds in length, made up of 6 coadditions of 10 seconds. To minimize 
overheads and permit timely focus adjustments due to truss temperature 
changes, exposures were run in 5 batches of 32 60-second exposures 
by automated scripts. To bridge the cosmetic defects on and the 
gap between detectors, each exposure was taken after random dithering within a 1$'$ x 1$'$ square.

The initial 99 pointings were observed for at least 160 minutes, with a median of 166 minutes of exposure time per pointing.  
We observed longer in some pointings (up to $\sim$400 minutes) when the seeing was poor, which was either noticed in real time at the telescope or at the end of each year when the stacks did not achieving the nominal depth.
The 10 supplemental pointings were 
observed for between 64 and 132 minutes with pairs of supplemental pointings 
overlapping by $\sim$60\% (to better overlap the DECam footprint). The median and standard deviation of the exposure time of non-suplemental pointings are 154 minutes and 45 minutes, respectively, and the distribution of the exposure times for all tiles is plotted in the right panel of Figure \ref{Fig:tile_stats}.
Some frames were not usable and omitted
due to poor seeing conditions.

\section{Image Reduction} \label{sec:reduction}
\subsection{Data Reduction and Calibration}
The NEWFIRM exposures were processed by the NOAO NEWFIRM calibration
pipeline \citep{swaters09} and were downloaded from the NOAO Science 
Archive\footnote{\url{http://archive.noao.edu/}}.  A detailed description of the NEWFIRM calibration pipeline    
reduction procedure can be found in the NOAO Data Handbook on the       
NOAO website\footnote{\url{http://ast.noao.edu/data/docs}}. We provide a     
brief summary of the procedure here. The NEWFIRM images were    
initially calibrated using exposures taken during the observing runs. The      
main calibration steps included artifact flagging, dark correction,     
linearity correction, flat-fielding, and sky subtraction. Next, the     
images were astrometrically calibrated with 2MASS reference images. 
Finally, the frames were remapped to a grid where each pixel is a       
square with a side length of 0.4$''$. The final data products included the remapped science images and their 
associated data quality maps (DQMs).

We inspected the quality and usefulness of every individual exposure 
with special attention given to the images       
flagged in the observers log as potentially flawed. Flawed images were removed 
because of artifacts provided by dew on the filter-housing window, 
elongated sources due to loss of the guide star or wind shake of the telescope truss,
triangular PSFs caused by loss of air pressure in the mirror supporting airbags, 
irregular structure in the background, and the erroneous masking of 
bright stars by the cosmic ray detection algorithm in the      
calibration pipeline.

During inspection, some pointings were found to have a number 
of exposures with corners exhibiting pixel values well above the typical 
background value. To avoid excluding
the entire image, we flagged by eye all the pixels in the ``bright corners'' of a maximum-combined pointing image aligned in 
pixel-space and flagged those pixels in each affected image's DQMs. 
This was repeated for each pointing as necessary. The pointings with 
masked ``bright corners'' include:
A8, A11, A25, B17, C14, C16, D6, and D10.3.

\subsection{Image Stacking}\label{subsec:stacking}
The remapped images that passed the visual inspection were combined 
using the PyRAF\footnote{\url{http://www.stsci.edu/institute/software\_hardware/pyraf}} imcombine function and a weighted mean procedure       
optimized for point-sources. The weighting of each image is a function of the seeing, transparency, and sky brightness and is defined by Equation A3 in \citet{gawiser06} as
\begin{equation}
w^{\mathrm{PS}}_{i} = \Bigg( \frac{\mathrm{factor}_i}{\mathrm{scale}_i \times \mathrm{rms}_i}\Bigg)^2,
\label{weightps}
\end{equation}

\noindent where $\mathrm{scale}_i$ is the image transparency (defined as the median brightness of the bright unsaturated stars after normalizing each star by its median brightness across all exposures), $\mathrm{rms}_i$ is the root mean square of the fluctuations of background pixels, and $\mathrm{factor}_i$ is defined as

\begin{equation}
\mathrm{factor}_i = 1 - \mathrm{exp} \Bigg( -1.3 \frac{\mathrm{FWHM}^{2}_{\mathrm{stack}}}{\mathrm{FWHM}^{2}_{i}} \Bigg),
\end{equation}

\noindent where $\mathrm{FWHM}_{\mathrm{stack}}$ is the median full-width at half maximum (FWHM) of bright unsaturated stars in an unweighted stacked image and $\mathrm{FWHM}_{i}$ is the median FWHM of bright unsaturated stars in each individual exposure.

The seeing and transparency measurements were determined using a preliminary source catalog generated for each resampled image using the \pkg{Source Extractor}\footnote{Version 2.25.0; \url{https://www.astromatic.net/software/sextractor}} software package \citep{bertin96}. The stars      
were selected from the stellar locus in the flux versus half-light      
radius parameter space and were required to appear in 80\% of the        
exposures per pointing. The aperture used to measure the brightness was one that collects 99\% of a star's light as determined by the        
curve of growth technique.
We quantified the noise
in each exposure by calculating the standard deviation of flux measured in 2$''$-diameter apertures randomly positioned across the image. When doing this, we used only the negative side of the distribution to avoid biasing the measurement with signal from real sources.

\begin{figure}[!tbp]
\includegraphics[scale=0.47,angle=0]{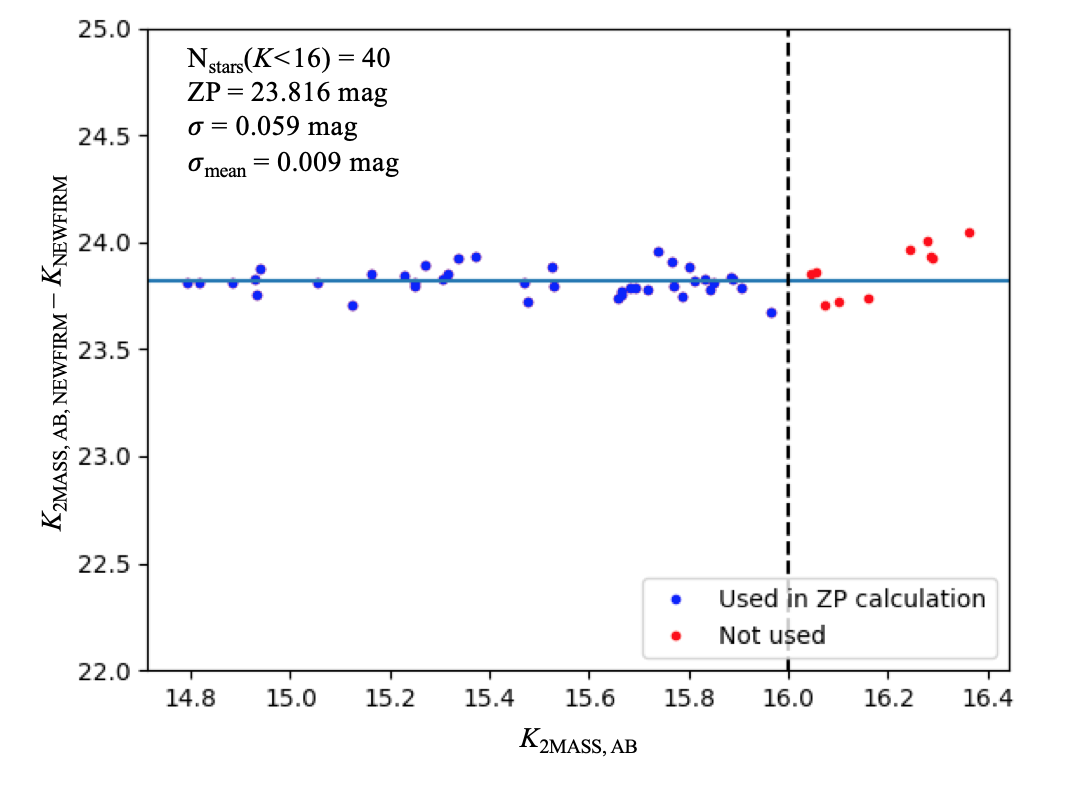}
\caption{Zero-point offset between 2MASS $K_{s}$ and NHS $K_{s}$ for bright stars in stacked image tile D1. Each tile has many bright unsaturated stars that were used to determine the zero-point offset.}
\label{Fig:zp}
\end{figure}

To flux calibrate, we compared      
aperture-corrected NEWFIRM instrumental magnitude to the $K_s$-band magnitudes in the 2MASS catalog. The aperture-corrected NEWFIRM instrumental magnitude were extracted using an aperture enclosing 70\% of the light from a point-source. This aperture size was determined using the curve of growth technique on a stacked image of 50 stars in the final image.
To convert 2MASS Vega magnitudes to AB, we start with
\begin{equation}
\text{m}_{\textrm{X}}-\text{m}_{\textrm{X,ref}}=-2.5\log{\frac{\int d\nu T_{\textrm{X}}(\nu)S_{\nu}/h\nu}{\int d\nu T_{\textrm{X}}(\nu)S_{\nu}(\textrm{ref})/h\nu}},
\label{eq2}
\end{equation}
where $T_{X}(\nu)$ is the X filter transmission function and $S_{\nu}$ is the source function. When converting from Vega to AB the reference source function is a flat spectrum with $S_{\nu}(\textrm{ref})\text{=3600 Jy}$, such that
\begin{equation}
\text{m}_{X}=\text{m}_{AB}-2.5\log{\frac{\int d\nu T_{X}(\nu)S_{\nu}/h\nu}{\int d\nu T_{X}(\nu)/h\nu}},
\label{eq3}
\end{equation}
where
\begin{equation}
\text{m}_{AB} = -2.5 \log (3600\ \textrm{Jy}) = -48.61\ \text{mag}
\label{eq4}
\end{equation}
when $S_{\nu}$ is in units of erg s$^{-1}$ cm$^{-2}$ Hz$^{-1}$. With $\text{m}_{X,Vega}\equiv0$, the Vega to AB magnitude offset for 2MASS $K_{s}$ is 1.8624 mag.
To put the 2MASS K AB magnitudes into the NEWFIRM $K_{s}$ magnitude system, we calculated the AB-Vega magnitude offset with the NEWFIRM $K_s$ filter (1.8763 mag) and added the difference between that and the same quantity for the 2MASS $K_s$ filter (0.0138 mag) to the 2MASS $K_{s}$ AB magnitudes.  This provided us with 2MASS measured magnitudes converted to both the AB system and the NEWFIRM $K_s$ filter. The zeropoint offset was calculated as the sigma-clipped (3-$\sigma$, 2 iterations) average difference for a statistically robust sample (N$>$20) of the brightest stars (typically $K_s<16-17$ mag) selected from the stellar locus of the flux versus half-light radius plot per stacked image. The zeropoint magnitude calculated for tile D1 is shown in Figure \ref{Fig:zp} as an example.

\begin{figure*}[!tbp]
\centering
\includegraphics[scale=0.4,angle=0]{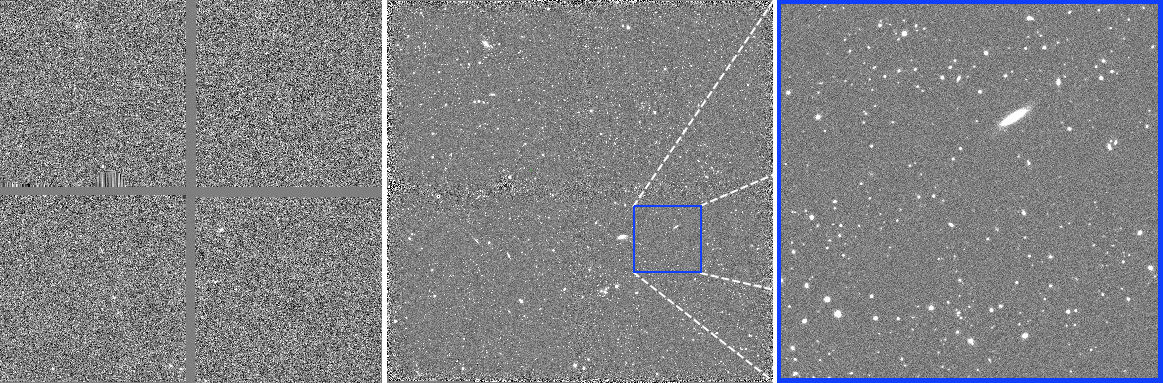}
\caption{Left: A single one-minute NEWFIRM exposure of tile B14 in the NHS. Center: The stacked image of tile B14. Right: Zoom-in $\sim$5\arcmin \ x $\sim$5\arcmin cutout of tile B14 stacked image. The stacked images are a combination of dithered single exposures and therefore have relatively smooth backgrounds, with data filling the cross-shaped detector gap seen in the single exposure. Each field overlaps neighboring tiles by at least 1\arcmin.}
\label{Fig:tile}
\end{figure*}

\begin{figure*}[!tbp]
\centering
\includegraphics[scale=0.7,angle=0]{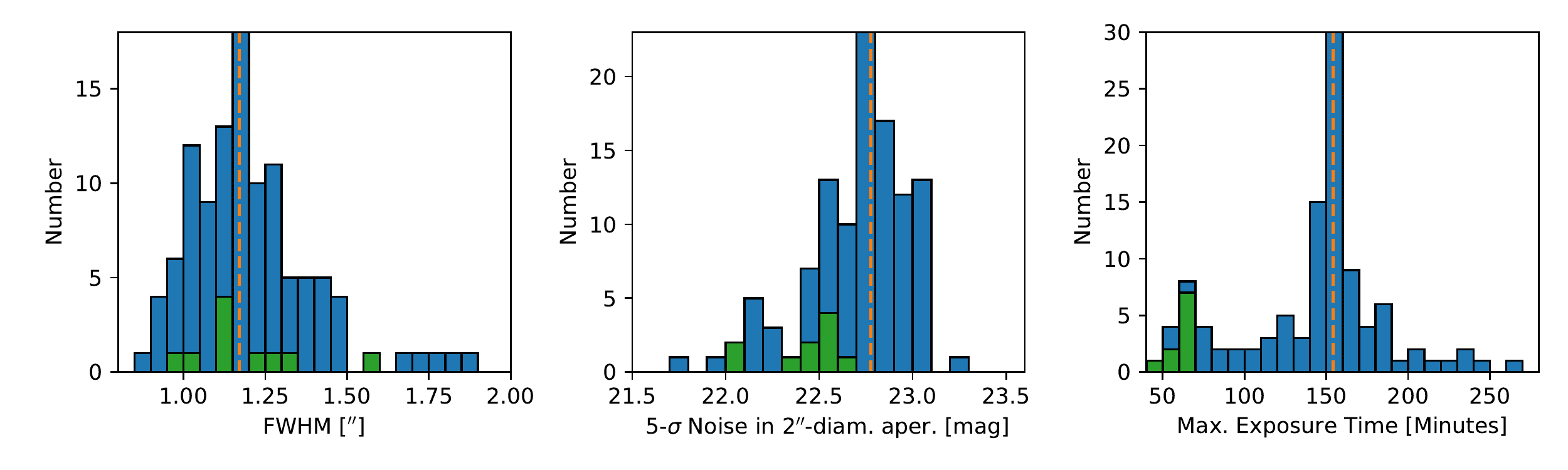}
\caption{The median FWHM or seeing (left), 5-$\sigma$ sky noise in $\sim$2$''$-diameter apertures as derived in Section \ref{subsec:stacking} (center), and maximum exposure time (right) for each NHS tile (blue). The distribution of the 10 supplemental tiles are highlighted in green. The median of each distribution is plotted as a vertical orange dashed line. The 5-$\sigma$ sky noise in 2$''$-diameter apertures for all fields relate to the 5-$\sigma$ depths for total fluxes by an added median aperture correction of -0.35 mag and a standard deviation of 0.12 mag, and these values are within $\lesssim$0.1 mag of those derived via the error functions in \S 4.2.}
\label{Fig:tile_stats}
\end{figure*}

The final NHS stacked images will be hosted on the NOAO Survey Archive\footnote{\url{https://www.noao.edu/survey-archives/}}.  As an example of our stacking process, in Figure \ref{Fig:tile} we show an individual image and the stack for tile B14. The distribution of stacked image qualities (seeing FWHM), 5-$\sigma$ depths for 2$''$-diameter apertures, and the maximum exposure times are plotted in the three panels of Figure \ref{Fig:tile_stats}. The median (standard deviation) for all 109 tiles are 1.17$''$ (0.19$''$), 22.78 mag (0.26 mag), and 154 minutes (45 minutes), respectively.  \edit1{These 2$''$-diameter aperture 5-$\sigma$ depths can be converted to the total depth for a point source by applying a median aperture correction of $-$0.35 mag.}

\subsection{Modeling Position-Dependent Point Spread Functions} \label{subsec:psfex}

The stacked images have non-uniform PSFs due to factors such as evolving seeing conditions and changes in telescope focus. However, within individual stacked images, the PSF varies smoothly as a function of position on the image. We modeled this behavior to obtain the most accurate aperture corrections, which allowed us to measure the highest-S/N fixed-aperture fluxes during the source cataloging procedure described in Section \ref{sec:cataloging}. 

We modeled the spatially variable PSF in each stacked image using the \pkg{PSF Extractor}\footnote{Version 3.18.2; \url{http://www.astromatic.net/software/psfex}} (\pkg{PSFEx}) software package \citep{bertin11}. This software is well documented and we follow the default procedure described in the documentation which we summarize here. 

The software takes a catalog from \pkg{Source Extractor} that includes the following for each detection: a small sub-image or ``vignette," centroid coordinate, half-light radius, signal-to-noise ratio in a Gaussian window, flux measured through a fixed aperture, flux uncertainty, object elongation, and extraction flags. Before running \pkg{Source Extractor} to create the input catalog, an rms image for each stacked image was needed for \pkg{Source Extractor} to generate reasonable flux uncertainties. We created this using the stacked science, DQM, and exposure time map images. The rms image was the inverse square of the exposure time map with the values scaled such that the median value of the resulting rms map equaled the sigma-clipped (3-$\sigma$, 2 iterations) standard deviation of pixel values in the science image where the exposure time is equal to the median and pixels contain no flux from objects. Objects are located using a segmentation map from a preliminary run of \pkg{Source Extractor} on each stacked image.

The \pkg{PSFEx} input catalog was created by running \pkg{Source Extractor} on each stacked image twice. The first run determined the pixel saturation level by finding the maximum pixel value of the brightest unsaturated star. The brightest unsaturated star was visually selected from the stellar locus on the flux versus half-light radius plot. We then set the gain and \kword{SATUR\_LEVEL} keywords to produce the binary fits catalog to be passed to \pkg{PSFEx}. The image gain was calculated as the average exposure time per pixel multiplied by the instrumental gain (8 electrons per ADU). The vignette size was set to 45 pixels by 45 pixels and the following keywords were set to the values in the parentheses for this exercise: \kword{DETECT\_MINAREA} (6), \kword{DETECT\_THRESH} (1.3), \kword{ANALYSIS\_THRESH} (1.3), \kword{DEBLEND\_NTHRESH} (32), \kword{DEBLEND\_MINCONT} (0.005), \kword{CLEAN} (\kword{Y}), and \kword{CLEAN\_PARAM} (1.0). 

Finally, we ran \pkg{PSFEx} on each stacked image with \kword{BASIS\_TYPE} = \kword{PIXEL} which models the PSF in pixel space as a linear combination of a limited number of known basis functions. We use \kword{BASIS\_NUMBER} = 31, indicating only the central 31 pixel by 31 pixel region of each source vignette was used in the modeling. A 30-pixel-diameter aperture was used for normalizing the amplitude of the PSF model. We ran \pkg{PSFEx} with a range of polynomials (n=2-8) and found the order at which the $\chi^2$ goodness of fit was minimized for each stacked image independently. We list the non-default parameter values used in the \pkg{PSFEx} runs in Table \ref{tab:psfex}.  These PSF models are used below to determine aperture corrections for fixed-aperture photometry.

\begin{deluxetable}{cc}[t!]
\tablecaption{Non-default \pkg{PSFEx} configuration parameters used \label{tab:psfex}}
\tablecolumns{2}
\tablewidth{0pt}
\tablehead{
\colhead{Parameter} &
\colhead{Value}  
}
\startdata
\kword{BASIS\_TYPE} & \kword{PIXEL} \\
\kword{BASIS\_NUMBER} & 31 \\
\kword{PSF\_SAMPLING} & 1 \\
\kword{PSF\_SIZE} & 31,31 \\
\enddata
\end{deluxetable}

\section{NHS Source Catalog} \label{sec:cataloging}
We detected and extracted sources by running \pkg{Source Extractor}\footnote{Version 2.25.0; \url{https://www.astromatic.net/software/sextractor}} on each NEWFIRM $K_s$ tile in one image mode with default \pkg{Source Extractor} parameters. The values of important or non-default \pkg{Source Extractor} parameters used are listed in Table \ref{tab:sextractor}. We selected the optimal values of \pkg{Source Extractor} parameters \kword{DETECT\_THRESH} and \kword{MINAREA} by considering the recovery fraction of simulated sources injected into the stacked images and the false positive fraction. This process is described in Section \ref{subsec:survey_depth}. The source flux errors were estimated empirically (Section \ref{subsec:errors}) and were used to find the aperture size that maximizes the S/N in each stacked image. Aperture corrections for the optimal apertures were determined using the position dependent PSF models (Section \ref{subsec:opt_apt}). Because the stacked images overlap on the sky, we combined the fluxes of sources in the overlap regions with a weighted mean in our final catalog (Section \ref{subsec:combine}).

\begin{deluxetable}{cc}[]
\tablecaption{Relevant \pkg{Source Extractor} configuration parameters used \label{tab:sextractor}}
\tablecolumns{2}
\tablewidth{0pt}
\tablehead{
\colhead{Parameter} &
\colhead{Value}  
}
\startdata
\kword{DETECT\_MINAREA} & 5 \\
\kword{DETECT\_THRESH} & 1.0 \\
\kword{ANALYSIS\_THRESH} & 1.0 \\
\kword{DEBLEND\_NTHRESH} & 32 \\
\kword{DEBLEND\_MINCONT} & 0.005 \\
\kword{CLEAN} & \kword{Y} \\
\kword{CLEAN\_PARAM} & 1.0\\ 
\kword{BACK\_SIZE} & 64 \\
\kword{BACK\_FILTERSIZE} & 3 \\
\enddata
\end{deluxetable}

\subsection{\edit1{Deriving Source Selection Parameters}} \label{subsec:survey_depth}
We selected the optimal values of \pkg{Source Extractor} parameters \kword{DETECT\_THRESH} and \kword{MINAREA} for our stacked images by injecting mock point-sources into our images and calculating the fraction that were recovered by \pkg{Source Extractor}. We inserted 50,000 mock point-sources into each stacked image using the modeled PSF from \pkg{PSFEx} (Section~\ref{subsec:psfex}) at random positions (excluding locations inside the isophotes of bright objects by a conservative amount). Mock sources were assigned magnitudes chosen randomly from a lognormal distribution in the range $K_{s}=$14.75-24.5 mag.

The recovery fraction of each magnitude bin was calculated as the number of mock sources recovered divided by the total number of mock sources injected. The false positive fraction per magnitude bin was calculated by multiplying the image by $-$1, re-running \pkg{Source Extractor}, and dividing the number of sources detected in this inverted image \edit1{(which are all noise spikes)} by the number of sources in the original stacked image.

\begin{figure}[!tbp]
\includegraphics[scale=0.3,angle=0]{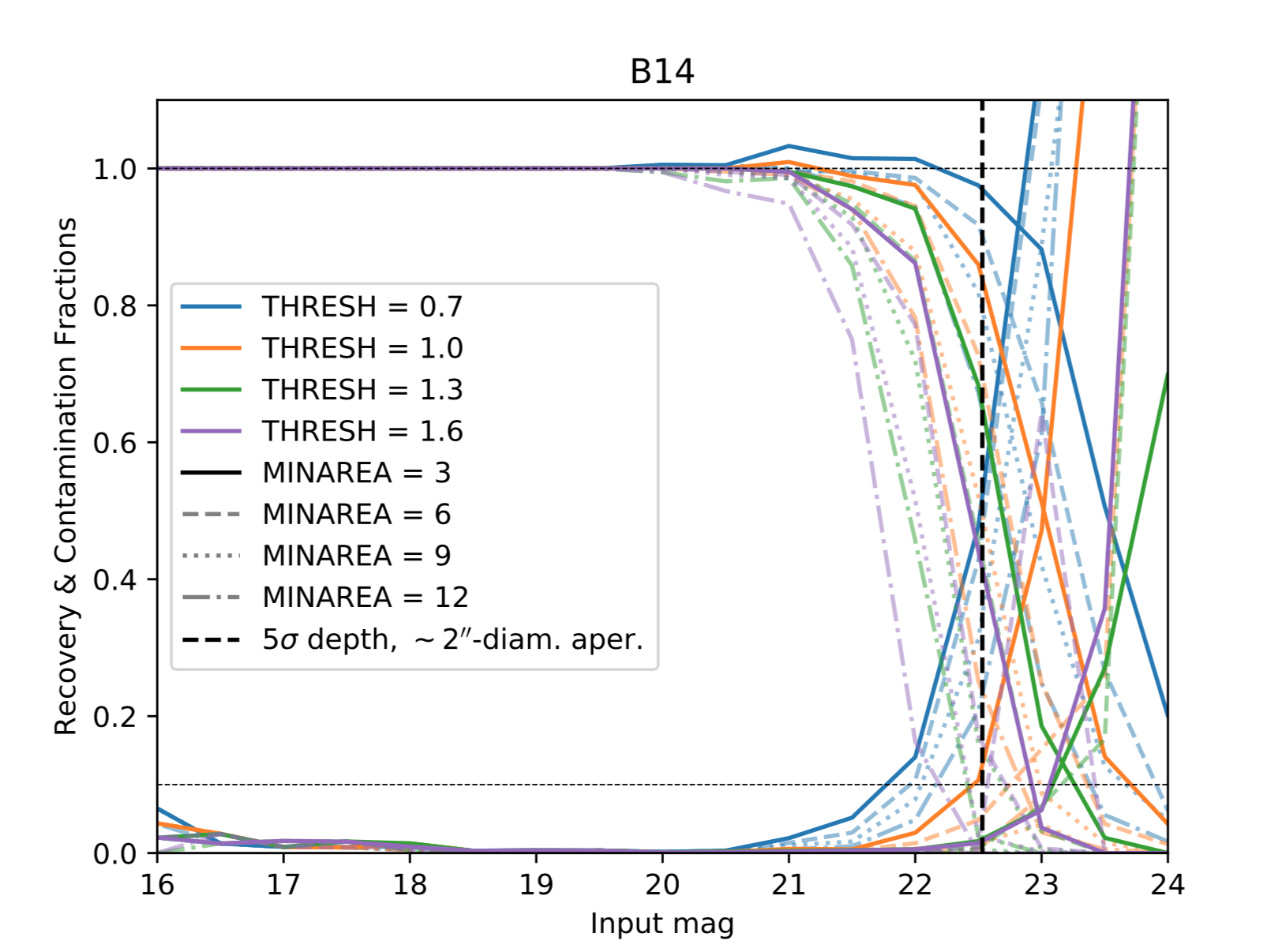}
\caption{Recovery and false-positive (or contamination) fractions for various combinations of \pkg{Source Extractor} selection parameters (see legend) in tile B14. The recovery fraction can be greater than 1.0 due to spurious sources. We selected the combination that had a false-positive fraction below 10\% and a recovery fraction above 50\% in the faintest magnitude bin for most of the stacked images. The selected combination was \kword{DETECT\_THRESH} = 1.0 and \kword{MINAREA} = 5. We used these parameter values for all stacked images for consistency.}
\label{Fig:recovery}
\end{figure}

The recovery and false-positive fractions were calculated from a number of \pkg{Source Extractor} catalogs generated using combinations of \kword{DETECT\_THRESH} and \kword{MINAREA} in the range 0.7--1.6 and 3--12, respectively. Figure \ref{Fig:recovery} shows the recovery and false positive fractions for field B14. We selected the combination of parameters that produced a false-positive fraction below 10\% and a recovery fraction above 50\% in the faintest magnitude bin for most of the stacked images. This analysis led to the final selected combination of \kword{DETECT\_THRESH} = 1.0 and \kword{MINAREA} = 5, which we used for all stacked images.

\subsection{Error Estimates} \label{subsec:errors}
\begin{figure}[!t]
\includegraphics[scale=0.6,angle=0]{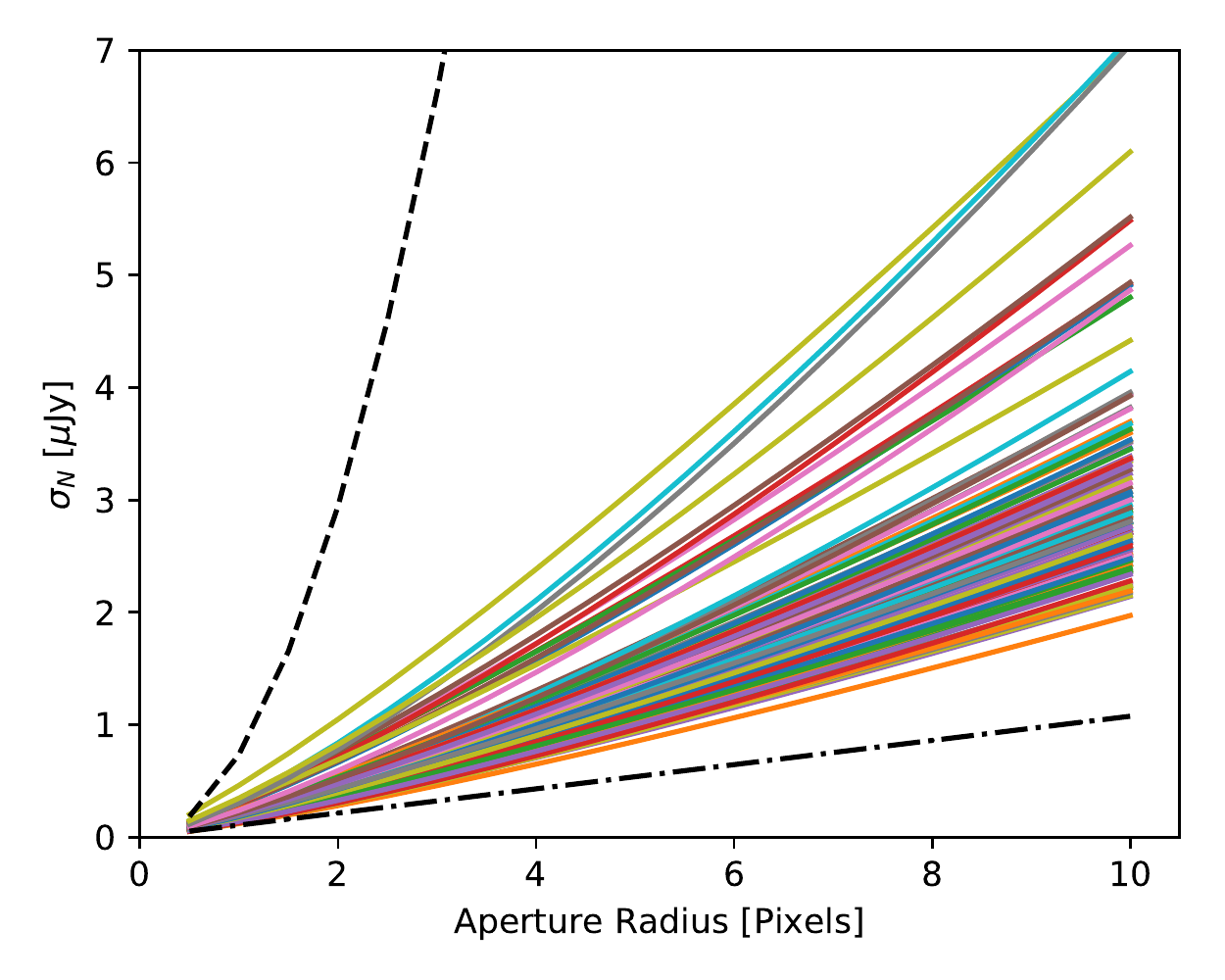}
\caption{Parameterized fits to the background noise fluctuations, $\sigma_{N}$, in an aperture of $N$ pixels plotted as a function of the aperture radius (in pixels) for all NHS tiles (colored lines). The dot-dashed black line shows the relation assuming uncorrelated pixels, $\sigma_{N} \sim \sqrt{N}$, for the tile with the largest $\sigma_{1}$. The dashed black line shows the relation assuming perfectly correlated pixels, $\sigma_{N} \sim N$, for the tile with the smallest $\sigma_{1}$ (\citealp{quadri07}). Our fits are in the partially correlated regime. We use these parameterized fits to estimate source flux uncertainties in the NHS catalog.}
\label{Fig:error}
\end{figure}

We estimated photometric uncertainties in the stacked images by fitting the image noise in apertures as a function of pixels per aperture, $N$, following the procedure described in Section 2.4 of \citet{stevans18} and based on the procedure of \citet{papovich16} \citep[see also][]{labbe03,gawiser06,quadri07,whitaker11}. Our procedure differs in the following ways. We randomly positioned 50,000 non-overlapping, circular apertures on unflagged pixels absent of source flux (i.e., avoiding source footprints in the segmentation maps from \pkg{Source Extractor}). We fit a parameterized function to the noise in an aperture of $N$ pixels, $\sigma_N$, with only two free parameters as,
\begin{equation}
\sigma_{N} = \sigma_{1} \alpha N^{\beta}
\end{equation}
\noindent where $\sigma_1$ is the pixel-to-pixel standard deviation in the sky background, and $\alpha$ and $\beta$ are free parameters. The parameterized fits to the background noise fluctuations functions are shown in Figure \ref{Fig:error}.

\subsection{Optimal-Aperture Sizes for Aperture Photometry of Point-Sources}\label{subsec:opt_apt}
\begin{figure}[!t]
\includegraphics[scale=0.85,angle=0]{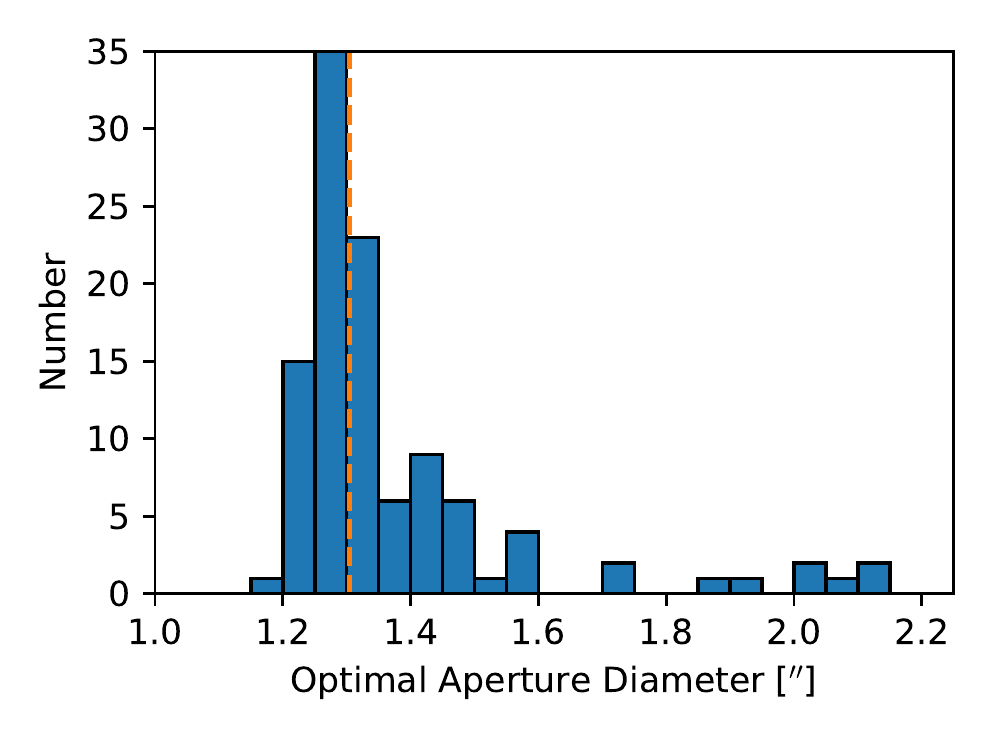}
\caption{Distribution of optimal-aperture diameters for all NHS stacked tiles. The median optimal-aperture diameter was 1.3$''$ with a standard deviation of 0.20$''$. These apertures produce the highest S/N flux measurements for point-sources.}
\label{Fig:opt_aper}
\end{figure}

After measuring each stacked image's unique noise properties, we found the optimal-aperture size for each stacked image by finding the aperture size that maximized the S/N for point-sources. We plotted the aperture S/N as a function of aperture size for sources in each magnitude bin, noted the aperture size with the highest S/N, and selected the modal value \edit2{giving a single optimal-aperture size per field}. The median optimal-aperture diameter of all stacked images was 1.3$''$ with a standard deviation of 0.20$''$. The distribution of optimal-aperture diameter is plotted in Figure \ref{Fig:opt_aper}.  We used the PSF model to calculate the localized aperture corrections for the optimal-aperture size of each stacked image \edit2{on a per-object basis}. The aperture corrections were calculated using the curve of growth technique on the model PSFs (Section \ref{subsec:psfex}) at 64 evenly spaced positions in a square grid across each tile.

\subsection{\edit1{AUTO Fluxes}}
\edit1{Any photometric measurement made using the optimal aperture, even after correcting to total, is appropriate only for a point source.  As the user of this catalog may be interested in resolved galaxies, we also include the Source Extractor ``AUTO'' flux.  This flux is measured in an elliptical Kron aperture, where we use the default parameters which are tuned to recover approximately the total flux.  We calculate uncertainties on these fluxes using the error functions derived in \S4.2 and the area covered by these elliptical apertures (calculated for each source using the Source Extractor returned A\_IMAGE, B\_IMAGE and KRON\_RADIUS values).  Both the fluxes and uncertainties are corrected to total using an aperture correction of 1/0.94, as the Source Extractor documentation shows that the default MAG\_AUTO apertures typically contain 94\% of the total flux.}

\subsection{Combining Catalogs of Overlapping Stacked Images} \label{subsec:combine}
Since the stacked images overlap their neighboring tiles by 1-2\arcmin \ on all sides, some sources will appear in 2-4 stacked images and their respective catalogs. We combined duplicate sources for the final catalog. Any source within 0.6$^{\prime\prime}$ of a source in another catalog was combined with each entry weighted by the inverse of its squared empirical error.

\begin{figure}[!t]
\includegraphics[scale=0.6,angle=0]{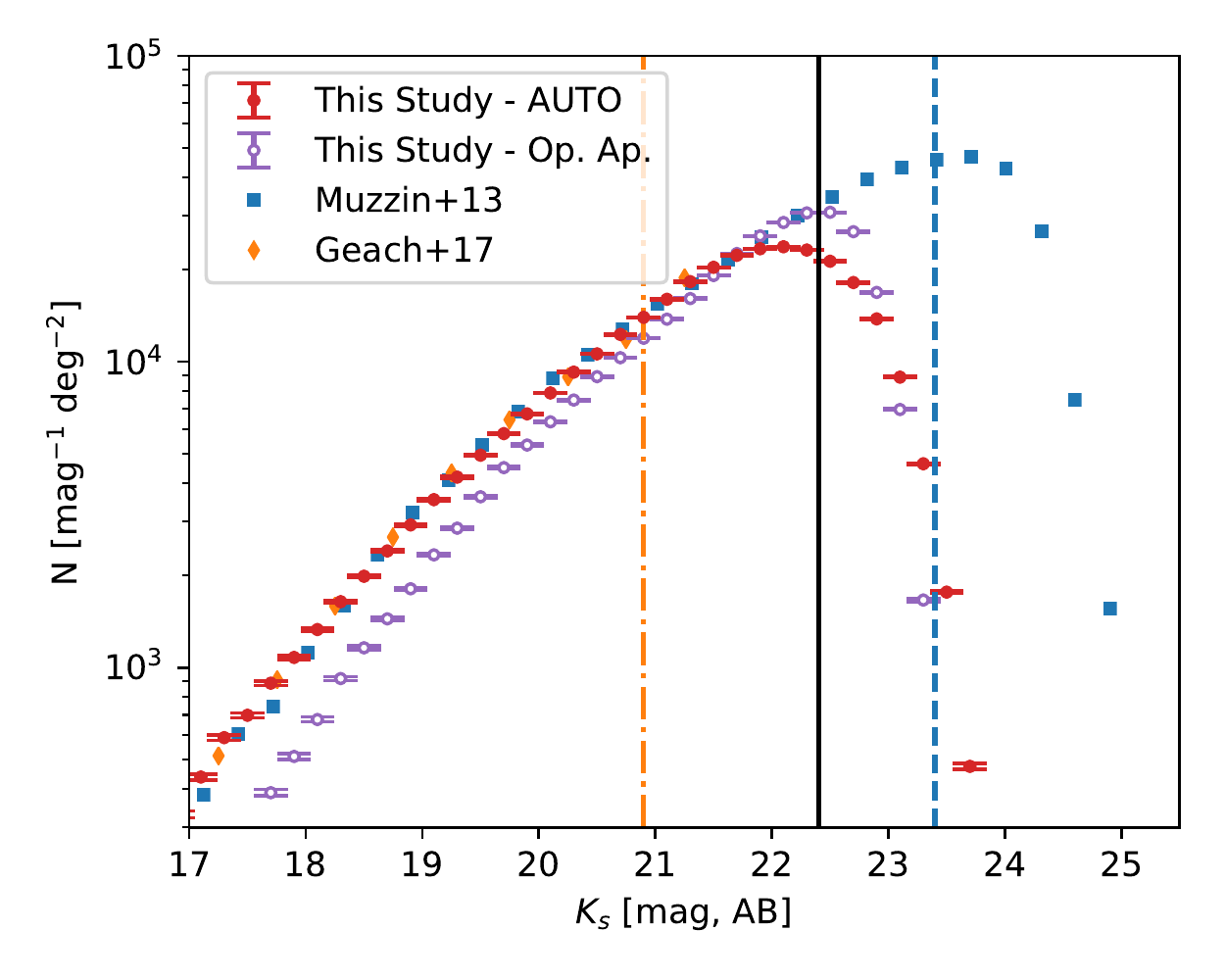}
\caption{\edit1{The NHS $K$-band galaxy differential number counts (with Poisson error bars)} compared to those measured in the deeper near-infrared survey UltraVISTA \citep{muzzin13} and the wider near-infrared survey VICS82 \citep{geach17} (see legend). \edit1{We include number counts derived using both \pkg{Source Extractor} AUTO apertures (filled purple circles) and from optimal apertures corrected to total (open red circles; see Section \ref{subsec:opt_apt}).} The NHS data have not been corrected for incompleteness. \edit1{The 5-$\sigma$ depth derived from 2$''$-diameter aperture measurements, corrected to total ($K_s=22.4$ mag, after a $-$0.35 mag aperture correction)}, is plotted as the black solid line. 
Number counts from the literature are reproduced from the published figures and the reported 5-$\sigma$ limits from UltraVISTA and VICS82 are shown as the dashed and dot-dashed lines, respectively. 
Comparing to the deeper UltraVISTA catalog, we find that the NHS number counts are 90\% complete at 22.0 (22.5) mag for AUTO (corrected optimal) apertures.  \edit1{Brighter than this, the NHS galaxy AUTO number counts agree well with both UltraVISTA and VICS82, while the corrected optimal circular apertures show a progressively larger shift towards fainter magnitudes as those apertures assume a point-source profile when correcting to total (while such bright sources will predominanly be resolved).}}
\label{Fig:n_counts}
\end{figure}

\subsection{Number Counts}
\edit1{We compare the number counts of galaxies in our $K_{s}$-band NHS catalog using both \pkg{Source Extractor} AUTO (apertures and optimally sized apertures (both corrected-to-total)} to values from the literature in Figure \ref{Fig:n_counts}. \edit1{This figure uses the final combined catalog, and thus effectively averages over differences in data quality between fields.}. \edit1{Galaxies are defined as sources with a  \pkg{Source Extractor} \kword{CLASS\_STAR} value of less than 0.95.} Number counts from the literature are extracted from the published figures. We find our number counts to be in excellent agreement with the deeper near-infrared survey UltraVISTA \citep{mccracken12} which covers 1.6 deg$^2$ to a depth of 23.8 mag \citep{muzzin13}. Our number counts are also consistent with the number counts from the VICS82 survey \citep{geach17} which covers $\sim150$ deg$^2$ to a depth of 20.9 mag. \edit2{This figure shows that our $K$-band selected catalog is $\gtrsim$90\% complete at $K \sim$ 22.0, and $\sim$70\% complete at our 5$\sigma$ limit of 22.4.}

\section{Matching to Ancillary Datasets} \label{sec:matching}

\subsection{IRAC Data Reduction and Photometry} \label{irac}
\par \edit1{In this subsection we discuss the photometric measurements of the SHELA {\it Spitzer}/IRAC imaging in this field.  Full details of this dataset are available in \citet{papovich16}, who find these data to be 50\% complete to a limiting magnitude of 22.6 at both 3.6 and 4.5 $\mu$m.}. To include these data in our catalog, we could position-match the published {\it Spitzer}/IRAC catalog from \citet{papovich16} to our NEWFIRM catalog, though this is not optimal for two reasons.  First, the \citet{papovich16} catalog is IRAC-detected, and so only includes sources with significant IRAC flux. For our purposes, even a non-detection in IRAC can be useful for calculating a photometric redshift.  Second, this catalog uses apertures defined by the positions and shapes of the IRAC sources. This is a significant limitation since the larger PSF of the IRAC data results in significant blending, especially at fainter magnitudes, where we expect to find the bulk of our high-$z$ sources.  For these reasons, we apply the \pkg{Tractor} image modeling code \citep{lang16a,lang16b} to perform  ``forced photometry,'' which employs prior measurements of source positions and surface brightness profiles from a high-resolution band to model and fit the fluxes of the source in the remaining bands, splitting the flux in overlapping objects into their respective sources. We specifically use \pkg{Tractor} to optimize the likelihood for the photometric properties of NEWFIRM $K_{s}$-band sources in each of the IRAC 3.6 and 4.5 $\mu$m bands given initial information on the source and image parameters.  The input parameters of the IRAC 3.6 and 4.5 $\mu$m images include a noise model, a PSF model, image astrometric calibration described by World Coordinate System (WCS), and calibration information (the ``sky noise'' or rms of the image background). The input source parameters include the NEWFIRM source positions, brightness, and surface brightness profile (i.e., effective radius, position angle, and axis ratio). \pkg{Tractor} then renders a model of a galaxy or a point-source, convolves it with the PSF model for each IRAC band, and performs a linear least-squares fit for source fluxes such that the sum of source fluxes is closest to the actual image pixels, with respect to the noise model.

We model the surface brightness profile of NEWFIRM $K_{s}$-band sources  using the same method as implemented in \citet{wold19}
for the IRAC forced photometry of DECam-selected sources. We refer the reader to \citeauthor{wold19}
for the full description of surface brightness profile modeling (Section 4.2) and the photometric error estimates (Section 4.3). Here we briefly summarize how we use \pkg{Tractor} to perform forced photometry on IRAC images for NEWFIRM $K_{s}$-band-selected sources below.

We begin with using the fluxes and surface brightness profile shape parameters measured in our NEWFIRM $K_{s}$-band image as our initial high-resolution model. We then use an empirical IRAC point-response functions (PRFs) for the 3.6 and 4.5 $\mu$m images (FWHM of $1\farcs97$ and $1\farcs99$ from Gaussian fits to the 3.6 and 4.5 $\mu$m PRFs, respectively) to model each source.  The construction of the IRAC PRF is described in \citet{papovich16} (Section 3.4). As in \citet{wold19}, we apply the scaling factor of 0.265 and 0.287 to 3.6 and 4.5 $\mu$m weighted images, respectively, to generate the rms maps for \pkg{Tractor} forced photometry procedure. These rms maps result in the median photometric error output from \pkg{Tractor} roughly matched with those measured in ${6}''$-diameter apertures, corrected to total, obtained from the published {\it Spitzer}/IRAC catalog \citep{papovich16}. 
For each NEWFIRM source, we extract an IRAC image cutout (${20}'' \times {20}''$) and measure its IRAC fluxes with three brightness profiles: a point-source profile, an exponential profile (equivalent to a S\'{e}rsic profile with $n=1$), and deVaucouleurs profile (equivalent to a S\'{e}rsic profile with $n=4$). The \pkg{Tractor} simultaneously modeled and optimized the sources of interest and neighboring sources within the cutout. Finally, the \pkg{Tractor} measured the IRAC flux of each NEWFIRM source with the lowest reduced chi-squared value, $\chi^2_{\mathrm{red}}$. In our final output catalog of IRAC forced photometry, we use IRAC fluxes based on the optimized model profile (those with the lowest $\chi^2_{\mathrm{red}}$ among the three brightness profiles).

\par We validated the \pkg{Tractor}-based IRAC fluxes by comparing the fluxes of isolated sources (no neighbors within  4$''$) to the published {\it Spitzer}/IRAC catalog from \citet{papovich16}. For both bands, we find good agreement with an offset of $\delta m = 0.05$ mag and a scatter of 0.15 mag down to $m=22$ mag.

\subsection{DECam Photometry} \label{subsec:decam}
\par \edit1{In addition to performing NEWFIRM-based forced photometry on the IRAC images, we also used \pkg{Tractor} to optimize the likelihood for the photometric properties of NEWFIRM $K_{s}$-band sources in all five DECam $ugriz$ bands.  The full details of these data are available in \citet{wold19}, which finds a typical image quality of $\sim$1.5$^{\prime\prime}$, and typical 5$\sigma$ limiting magnitudes of $u\sim$25.4, $g\sim$25.1, $r\sim$24.6, $i\sim$24.0 and $z\sim$23.7 from their PSF-matched data (see their Table 3 for full details).}  The input image parameters of the $ugriz$ images include the empirical DECam PSFs, astrometric calibration, and sky noise for the stacked DECam images from \citet{wold19}. We apply scaling factors to the DECam weighted images in each band and tile them to generate the rms map for the \pkg{Tractor} forced photometry procedure. We then used the same method as described in Section~\ref{irac} and \citet{wold19} to model the brightness profile and estimate photometric errors for all sources. In our final output catalog of DECam forced photometry, we quote DECam fluxes based on the optimized model profile.

We validate the \pkg{Tractor}-based DECam fluxes by comparing the fluxes of isolated sources (no neighbors within  4$''$) to the published $ugriz$-band DECam catalogs for the {\it Spitzer}/SHELA survey \citep{wold19}. For  $u', g', r', i',$ and $z'$ bands, we find good agreement with a median bias offset of $\delta m = [0.04,0.03,0.06,0.05,0.04]$ mag and a scatter of $[0.14,0.13,0.11,0.09,0.10]$ mag, respectively, down to $m=26$ mag \citep[see][for the FWHMs from Gaussian fits to the five $ugriz$-bands and four tiles]{wold19}.

\begin{figure*}[!t]
\centering
\includegraphics[scale=0.6,angle=0]{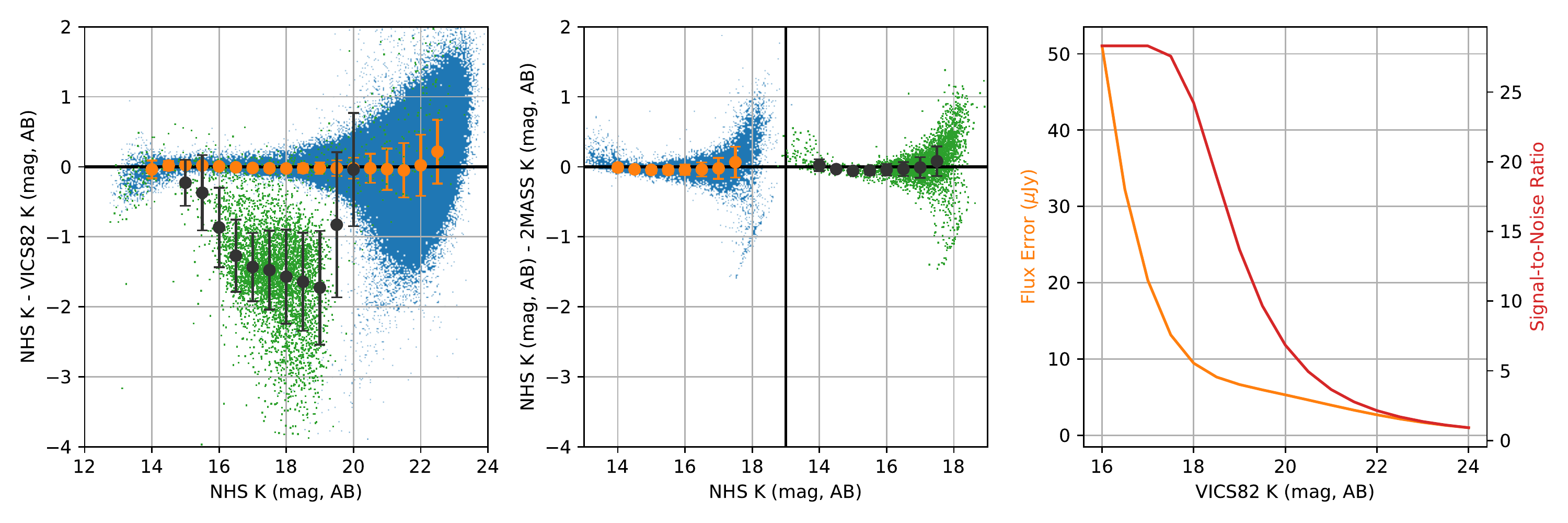}
\caption{Left: Magnitude difference between NHS NEWFIRM $K_{s}$ and VICS82 $K_{s}$ filters for point-sources (blue) with the binned sigma-clipped median and standard deviations (orange). A locus of sources with significant offsets is seen (green) along with their binned sigma-clipped median and standard deviations (black). A similar locus is seen when comparing VICS82 $J$ to 2MASS $J$. Center: Magnitude difference between NHS NEWFIRM $K_{s}$ and 2MASS $K_{s}$ filters for the same sources shown in the left panel plotted (when detected in 2MASS). The sources in the offset locus in the left panel show no offset between NEWFIRM $K_{s}$ and 2MASS $K_{s}$. No filter offsets were applied in these figures. In our catalog, we flagged VICS82 fluxes of sources in the offset locus by identifying objects with weight map values of zero within the central 3x3 pixels. Values from these objects are not included in our catalog.  Right: Our derived flux errors assigned to VICS82 $J$ and $K_{s}$ fluxes as a function of magnitude (orange line) along with the resulting signal-to-noise ratio (red line). The resulting 5-$\sigma$ depths, $J=K_{s}=20.75$ mag, are consistent with the depths reported for the VICS82 catalog ($J=21.3$ mag and $K_{s}=20.9$ mag) in \citet{geach17}. }
\label{Fig:vics82}
\end{figure*}

\subsection{Position-matching to VICS82}
The VISTA-CFHT Stripe 82 Near-infrared Survey\footnote{\url{http://stri-cluster.herts.ac.uk/vics82/}} (VICS82; \citealt{geach17}) covers $\sim$85\% of the optical imaging footprint of our field with $J=21.3$ mag and $K_{s}=20.9$ mag imaging. The $J$-band data adds a missing wavelength to our catalog, and the $K_{s}$-band photometry, while $\sim$2 magnitudes shallower than our NHS data, adds an independent measurement. We position-matched our source catalog to the publicly available February 2017 version of the VICS82 catalog \citep{geach17}. The matching radius used was 1.2$\arcsec$.

On inspection of the magnitude difference between VICS82 $K_{s}$ and 2MASS $K_{s}$ and between VICS82 $K_{s}$ and our NEWFIRM $K_{s}$ (Figure \ref{Fig:vics82}; left panel), we found a locus of sources with $\Delta$mag offsets $\sim$0.3-3 mag (green markers). A similar feature was found when comparing the VICS82 $J$ source magnitudes to that of 2MASS $J$. We inspected the VICS82 science and weight map images at the position of some of the offset sources and found weight map pixel values of zero near the object centers, which led to underestimated fluxes. We flagged and removed any VICS82 flux measurements where the minimum weight-map pixel value within the 3 x 3 pixels of object centers is zero in either $J$ or $K_{s}$. This resulted in the removal of 1\% of VICS82 sources with NHS counterparts.

The February 2017 version of the VICS82 catalog contained no source error estimates. An indirect error estimate was contained in the Gaussian-weighted S/N column, but these errors are inappropriate for aperture photometry. To estimate the uncertainties of VICS82 flux measurements we assigned errors based on the standard deviation of the difference between VICS82 and NHS magnitudes as a function of VICS82 $K_{s}$ magnitudes. We fitted the standard deviation from $K_{s}=$17-21.5 mag using the function $\alpha*$STDDEV$^{\beta}$, where $\alpha$ and $\beta$ are free parameters, after normalizing by the standard deviation in the mag=17 bin. For the $J$ band, we adopted the $K_{s}$-band error function because we did not have an independent $J$-band catalog of sufficient depth. The VICS82 catalog states that the $J$-band images are deeper than $K_{s}$, so our $J$ errors are conservatively large. The right panel of Figure \ref{Fig:vics82} shows the estimated flux errors assigned to the VICS82 $J$ and $K_{s}$ fluxes as a function of magnitude along with the resulting signal-to-noise ratio. The 5-$\sigma$ depths are $J=K_{s}=20.75$ mag, which are similar to the 5-$\sigma$ depths \edit1{reported for the VICS82 catalog for $K_{s} (20.9$), and lower for $J$ (21.3)}.

\edit1{We note that although we have used multiple photometric techniques (Source Extractor for $K_s$, Tractor for DECam and IRAC, and the VICS82 published catalog), we have strong evidence that there is no systematic bias in the measured photomtry due to these specific techniques.  For DECam and IRAC we find excellent agreement with previously published Source Extractor catalogs by \citet{wold19} and \citet{papovich16}, respectively, and with VICS82, with the exception of the small number of outliers, we see excellent agreement with our own $K_s$-band photometry.  One final piece of evidence comes from \citet{sherman20}, who used our catalog and derived photometric offsets calculated by finding the best-fit template for a set of galaxies with known spectroscopic redshifts with EAZY-py.  They measured the median offsets between those template fluxes and the measured fluxes, finding a maximum offset of <2\%, with $<$1\% in the majority of filters.}

\subsection{Correcting Photometry for Galactic Extinction}
All source fluxes and errors were corrected for Galactic extinction using the color excess $E(B-V)$ measurements by \citet{schlafly11} We obtained $E(B-V)$ values using the Galactic Dust Reddening and Extinction application on the NASA/IPAC Infrared Science Archive (IRAS) website\footnote{\url{http://irsa.ipac.caltech.edu/applications/DUST/}}. We queried IRAS for E(B-V) values (i.e., the mean value within a 5$\arcmin$ radius) for a grid of points across the SHELA field. The grid had 120 values in right ascension ranging from 13$\arcdeg$-25$\arcdeg$ and 50 values in declination ranging from -1$\fdg$25 to 1$\fdg$25. The median $E(B-V)$ was 0.0269$\pm$0.007 (the minimum and maximum were 0.0159 and 0.1004, respectively). The $E(B-V)$ value of the grid point closest to each source was assigned. The \citet{cardelli89} Milky Way reddening curve parameterized by $R_V=3.1$ was used to derive the corrections at each band's central wavelength.

\section{Galaxy Properties from \pkg{eazy-py}} \label{sec:eazypy}
We measured photometric redshifts and galaxy physical properties (e.g., stellar mass, star formation rate, rest-frame colors) by running the full NHS photometry catalog including ancillary data (Section \ref{sec:matching}) through the publicly available software package \pkg{eazy-py}\footnote{Version 0.2.0-16-g6ab4498; \url{https://github.com/gbrammer/eazy-py}} which is based on the \pkg{EAZY} code \citep{brammer08}. \pkg{eazy-py} finds the linear combination of 12 Flexible Stellar Population Synthesis (FSPS) templates \citep{Conroy09, Conroy10} that minimizes the $\chi^2$ with respect to the extinction-corrected fluxes in all available photometric bands (i.e., DECam, NEWFIRM, VISTA and IRAC). For \edit2{the remainder of our analysis, we used the Kron-based NEWFIRM flux from \pkg{Source Extractor} (e.g., \kword{FLUX\_AUTO}) corrected to total, such that these results should be appropriate even for resolved sources.  Fluxes in all other bands also represent total fluxes, as described above.}

The non-default configuration parameters we used to run \pkg{eazy-py} are listed in Table \ref{tab:eazypy}. We ran \pkg{eazy-py} with flat priors because we are exploring a volume and parameter space larger than existing surveys and simulations, which are often used to derive priors and can suffer from large uncertainties due to small number statistics and cosmic variance especially at the bright end of the luminosity function. This could bias redshift solutions for real high redshift massive galaxies in our survey towards lower redshifts. We used the recommended set of templates from `templates/fsps\_full/tweak\_fsps\_QSF\_12\_v3.param' (G.\ Brammer, private communication) which utilize a \citet{chabrier03} initial mass function, \citet{kriek13} dust law, and solar metallicity. The templates include a range of galaxy types (e.g., star-forming, quiescent, dusty) and realistic star formation histories (SFH; e.g., slowly rising, slowly falling, and bursty).

\begin{deluxetable}{cc}[!]
\tablecaption{Non-default \pkg{eazy-py} configuration parameters used \label{tab:eazypy}}
\tablecolumns{2}
\tablewidth{0pt}
\tablehead{
\colhead{Parameter} &
\colhead{Value}  
}
\startdata
\kword{Z\_MAX} & 12 \\
\kword{Z\_STEP} & 0.01 \\
\kword{SYS\_ERROR} & 0.1 \\
\kword{APPLY\_PRIOR} & False \\
\kword{CAT\_HAS\_EXTCORR} & False \\
\kword{MW\_EBV} & 0 \\
\kword{PRIOR\_FILTER} & 134 \\
\kword{PRIOR\_FILE} & `templates/prior\_K\_extend.dat' \\
\enddata
\end{deluxetable}

\begin{figure}[!tbp]
\centering
\includegraphics[scale=0.65,angle=0]{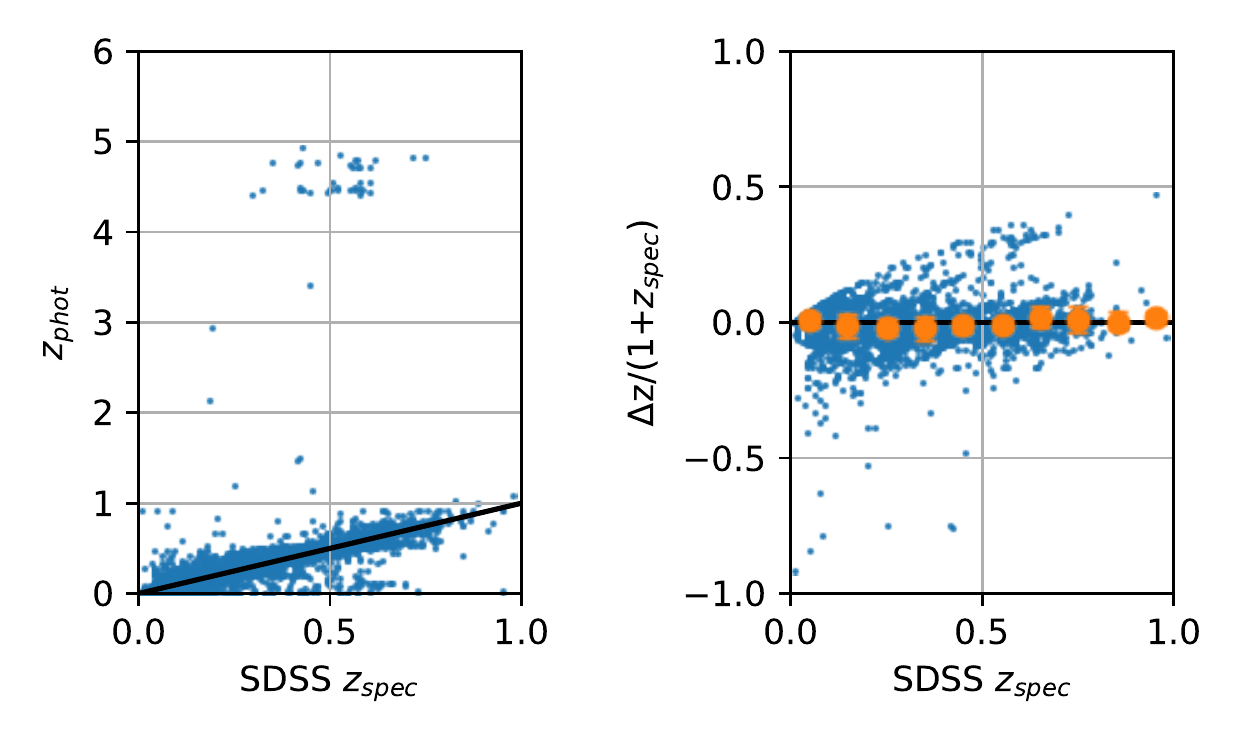}
\caption{Photometric redshifts compared to spectroscopic redshifts for 4,513 galaxies from SDSS DR12. We find good agreement for objects that have $K_{s}$ magnitudes in the $K_{s}$-magnitude range of our candidate $z=3-5$ galaxy sample ($K_{s}> 18.3$ mag) implying our sample is free from contamination from objects like those observed in SDSS. The 42 outliers in the upper part of the left panel are due to poor modeling of extended-source structures like spiral arms of bright galaxies, leading to incorrect optical fluxes (see discussion at the end of Section \ref{sec:eazypy}).}
\label{Fig:spec_z}
\end{figure}

To validate the photometric redshifts we compared them to 4,513 spectroscopic redshifts in the field from SDSS DR12 (Figure \ref{Fig:spec_z}), which were mostly at $z <$ 1. We found good agreement for objects without complex extended structure (e.g., spiral arms), for which our photometry extraction was optimized. We found a median offset of $\delta z / (1+z_{\textrm{spec}}) = 0.009$, a normalized median absolute standard deviation of $\sigma_{\textrm{NMAD}} = 0.04$, and a 5-$\sigma_{\textrm{NMAD}}$ outlier fraction of $<5$\%. 
A small population of 41 objects with $z_{\textrm{spec}}<1$ were measured to have $z_{\textrm{phot}}> 4$.  
These objects were brighter than our brightest candidate $z>3$ galaxies ($K_{s}<18.3$ mag)--none of the 730 fainter objects in the $z_{\textrm{spec}}$ sample were fit at $z_{\textrm{phot}}> 4$. It is unlikely that objects like these (selected for spectroscopic observations with SDSS) are contaminating our entire high-$z$ galaxy sample. Fainter objects have lower signal-to-noise ratios and therefore have more photometric scatter, so one would expect fainter objects to scatter to higher $z_{\textrm{phot}}$ more often if the population with $z_{\textrm{spec}}<1$ and $z_{\textrm{phot}}> 4$ was due to only random error. On inspection of the photometry of this outlier sample (with $z_{\textrm{spec}}<1$ and $z_{\textrm{phot}}> 4$), we saw complex structure like spiral arms and bars, which were not properly fit by the more simple light-profile models used with our photometry measuring code \pkg{Tractor} (See Section \ref{sec:matching}). Considering that the rest of the sample with $z_{\textrm{spec}}<1$--especially those with fainter fluxes ($K_{s}>18.3$ mag) and simpler structures--had measured $z_{\textrm{phot}}$ that were consistent with $z_{\textrm{spec}}$, we therefore conclude our values $z_{\textrm{phot}}$ are \edit1{as reliable as can be given the current spectroscopic constraints in this field, though future spectroscopic efforts can better validate this catalog at $z >$ 1.}

The electronic version of this paper will include a full machine-readable table of our full catalog, including both photometry and EAZY-py properties.  
A sample of 10 lines of the catalog is given in Table \ref{mcat1} the Appendix, and the full machine-readable table is available with the electronic version of this paper. The photometric redshifts provided in the table are determined by \pkg{eazy-py}, using the parameter ``z\_phot'' which fits a parabola to the array of $\chi^2$ with respect to redshift \edit2{across a three element window centered on the minimum $\chi^2$ value} and finds the minimum of the parabola.

\section{Massive Quiescent Galaxies at $z = 3-5$} \label{sec:results}

While today nearly all massive (log M/M\sol\ $>$ 11) galaxies are quiescent, forming few stars, at higher redshifts such galaxies continue to grow.  The frontier is now at $z >$ 3, where simulations find that nearly all such massive galaxies should be highly star-forming \citep[e.g.][]{brennan15}, with massive quiescent galaxies at $z >$ 4 being exceedingly rare.  Observations find a wide range of quiescent fractions amongst massive galaxies at $z \sim$ 3 \citep[e.g.][]{muzzin13,stefanon13}, which may be due to small fields covered by most studies.  Observations have yet to discover a significant population of massive quiescent galaxies at $z >$ 4. As an example of the utility of the NHS $K$-band selected catalog over $>$20 deg$^2$, here we demonstrate a conservative selection of massive quiescent galaxies at $z >$ 3.

\subsection{Selection of Massive Galaxies at 3 $< z <$ 5} \label{subsec:photoz}
We selected $z=3-5$ galaxies using criteria based on photometric cuts and properties of the photometric redshift ($z_{\mathrm{phot}}$) probability distribution functions ($z$PDFs) similar to the procedure outlined by \citet{fink15}. Prior to these EAZY-based cuts, we required a 5-$\sigma$ or greater detection significance in $K_{s}$, a 2-$\sigma$ or greater detection significance in IRAC 3.6 $\mu$m, and a $u'$-band S/N $<2$, as the Lyman break falls red-ward of the $u'$ band in galaxies at $z>3$. We also required a measurement (but not necessarily a detection) in all 5 optical DECam bands to ensure reliable constraints on $z_{\mathrm{phot}}$.

One limitation of our study is that red galaxies at low redshift can have SED shapes that appear similar to our $z=3-5$ galaxies within our filter set. We discussed this topic in the context of existing SDSS spectral observations in our field in Section \ref{sec:eazypy}. The degeneracy of these objects in optical and IR photometry was investigated by \citet{dunlop07} and motivated the inclusion of a red, dusty template in \pkg{eazy-py} \citep{brammer08} to account for the presence of these objects.  This results in several potential candidate high-redshift quiescent galaxies having small peaks in their P($z$) distributions at $z <$ 2.  Becasuse the abundance of red galaxies at $z <$ 2 is likely much higher than at $z >$ 2, this could potentially lead to high contamination rates.

\begin{deluxetable*}{cccccccccccc} 
\tablecaption{Candidate Massive Quiescent Galaxies} 
\tablehead{\colhead{ID} & \colhead{RA} & \colhead{Dec} & \colhead{$K_{AB}$} & \colhead{$z_{p}$} & \colhead{$\int P(z_{p}) >$ 3} & \colhead{log(M$_{\ast}$)} & \colhead{log(sSFR)} & \colhead{log(M$_{\ast}$)} & \colhead{log sSFR} & \colhead{$\int$P(sSFR)$<-$10.05}\\
\colhead{} & \colhead{(J2000)} & \colhead{(J2000)} & \colhead{} & \colhead{} & \colhead{} & \colhead{EAZY} & \colhead{EAZY} & \colhead{BC03} & \colhead{BC03} & \colhead{BC03}\\
\colhead{} & \colhead{(deg)} & \colhead{(deg)} & \colhead{(mag)} & \colhead{$ $} & \colhead{$ $} & \colhead{(M\sol)} & \colhead{(yr$^{-1}$)} & \colhead{(M\sol)} & \colhead{(yr$^{-1}$)} & \colhead{$ $}}
\startdata 
67411&14.996527&-0.347113&19.9&3.87&0.71&12.3&-11.4&12.3&-11.0&0.95\\
291064&16.754635&-0.028062&20.7&4.63&0.90&12.2&-10.2&12.2&$<$-11.5&0.98\\
311951&16.762005&-0.358180&20.6&4.63&0.95&12.4&-10.1&12.4&-10.3&0.74\\
344891&17.027281&-0.697423&20.4&3.43&0.84&11.9&-11.5&11.9&$<$-11.2&0.97\\
386425&17.379851&0.729148&20.3&3.99&0.76&11.7&-11.1&11.7&$<$-11.0&0.96\\
664819&19.731433&0.492264&21.0&4.65&0.89&11.7&-10.2&11.7&$<$-11.0&0.77\\
1055967&22.746300&0.370704&20.6&4.03&0.87&12.2&-11.6&12.0&-10.3&0.92\\
1198361&24.057372&0.135012&20.8&4.44&0.99&11.9&-10.3&11.8&-10.6&0.71\\
1208324&23.806994&-0.483853&19.6&4.15&1.00&12.0&-10.1&12.0&$<$-11.3&0.99
\enddata  
\tablecomments{Properties of the final sample of nine candidate massive quiescent galaxies.  $^{a}$Objects with SFR$_{BC03} <$ 5 M\sol\ yr$^{-1}$ have their sSFR listed as an upper limit derived with this limiting value (corresponding roughly to the UV-based SFR at the limit of the DECam optical imaging).}
\label{qgcat}
\end{deluxetable*}

One way to mitigate this is to use an apparent magnitude prior, which applies a Bayesian prior based on a source's apparent magnitude. EAZY has a built in such a prior, devised based on the luminosity functions from the \citet{delucia07} semi-analytic model (see \citealt{brammer08} for details).  These priors are typically peaked at low-redshift with a tail to higher redshift, with the tail becoming more prominent at fainter magnitudes.  As our goal here is simply to demonstrate the utility of the NHS catalog for discovering massive quiescent galaxies at $z >$ 3, we elect to make use of this prior when calculating photometric redshifts for this analysis to minimize contamination as much as possible.  For bright sources, the application of this prior will typically result in any low-redshift solutions becoming the dominant one, even if the higher-redshift solution was a better fit to the photometry.  We note that this likely results in significant incompleteness in our sample.  The simulation used for this prior used only a 1 deg$^2$ light cone, thus does not have the volume to constrain the abundance of the rare sources we seek here.  Additionally, as this simulation is $>$10 years old, it would not have made use of the advances in galaxy feedback physics made over the intervening years which have increased the efficiency of quenching in more modern simulations.  Nevertheless, here we include the prior to ensure a conservative selection, and we will revisit the prior in a future work.

We applied the following four selection requirements based on the EAZY $z_{\mathrm{phot}}$ and stellar mass results, using the results ran with the apparent magnitude prior: 
\begin{enumerate}
\item the highest peak of the $z$PDF must fall at $z >$ 3,
\item the integral of the normalized $z$PDF between $3 \leq z < 5$ must be greater than or equal to 0.6, which ensures a majority of the $z$PDF integrated probability is in the redshift range of interest,
\item the best-fitting template from \pkg{eazy-py} must have a $\chi^2$ $<$ 10
\item the best-fitting stellar mass $>$ 10$^{11}$ M\sol\
\end{enumerate} 

This sample of massive (log M/M\sol\ $>$ 11) high-redshift (3 $< z <$ 5) consists of 466 galaxy candidates from our full catalog.  We retierate that this is a lower limit on the true number, due to the use of the luminosity prior which increases the fidelity of our sample at the cost of increased completeness.

\subsection{Selecting Massive Quiescent Galaxies} \label{subsec:sel-qgals}
It is common to separate quiescent galaxies from star-forming galaxies using a selection box in the rest-frame $U-V$ vs $V-J$ diagram \citep[$UVJ$ diagram; e.g.,][]{labbe05, wuyts07,  whitaker11, muzzin13} given the bi-modality of galaxies in color and SFR \citep[e.g.,][]{ kauffmann03} even at high-redshift \citep[e.g.,][]{bell04, brammer09}. Another separation method used is setting a threshold in sSFR derived from SED-fitting, which has been found to correlate with $UVJ$ diagram classifications up to z = 2.5 \citep[e.g.,][]{williams10}. 

We select candidate quiescent galaxies from this parent massive galaxy sample using sSFR.  To decide on the appropriate sSFR threshold, one must consider the evolution in the galaxy main sequence, where galaxies at fixed stellar mass have higher average star-formation rates at higher redshift \citep[e.g.,][]{noeske07,whitaker14,salmon15}.  We use the relation found by \citet{salmon15} at $z =$ 4 to decide on a sSFR cut.  They find log (SFR/M\sol/yr) $=$ 0.7 log (M/M\sol) $-$ 5.7 using a sample of galaxies spanning 8.5 $<$ log (M/M\sol) $<$ 10.5, with a 1$\sigma$ scatter of 0.35 dex in SFR at fixed stellar mass.  Extrapolating mildly to our stellar mass threshold of log (M/M\sol) $=$ 11, this relation predicts the typical galaxy should have log (SFR/M\sol/yr) $=$ 100 M\sol\ yr$^{-1}$, for a log (sSFR/yr$^{-1}$) $=$ $-$9.  To classify a galaxy as quiescent, we require it to have a SFR which is 3$\sigma$ below this threshold.  This corresponds to a SFR $=$ 8.9 M\sol\ yr$^{-1}$, or log(sSFR/yr$^{-1}$) $<$ $-$10.05.  Applying this sSFR cut to the initial massive galaxy sample results in a sample of 51 candidate massive quiescent galaxies.

\begin{figure*}[!t]
\centering
\includegraphics[scale=0.6,angle=0]{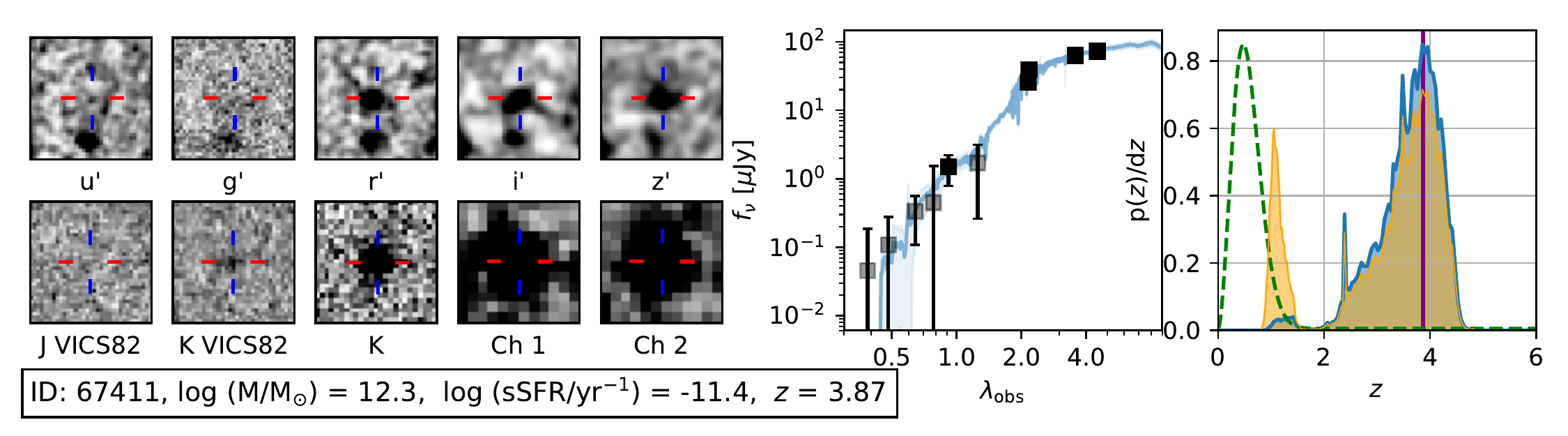}
\includegraphics[scale=0.6,angle=0]{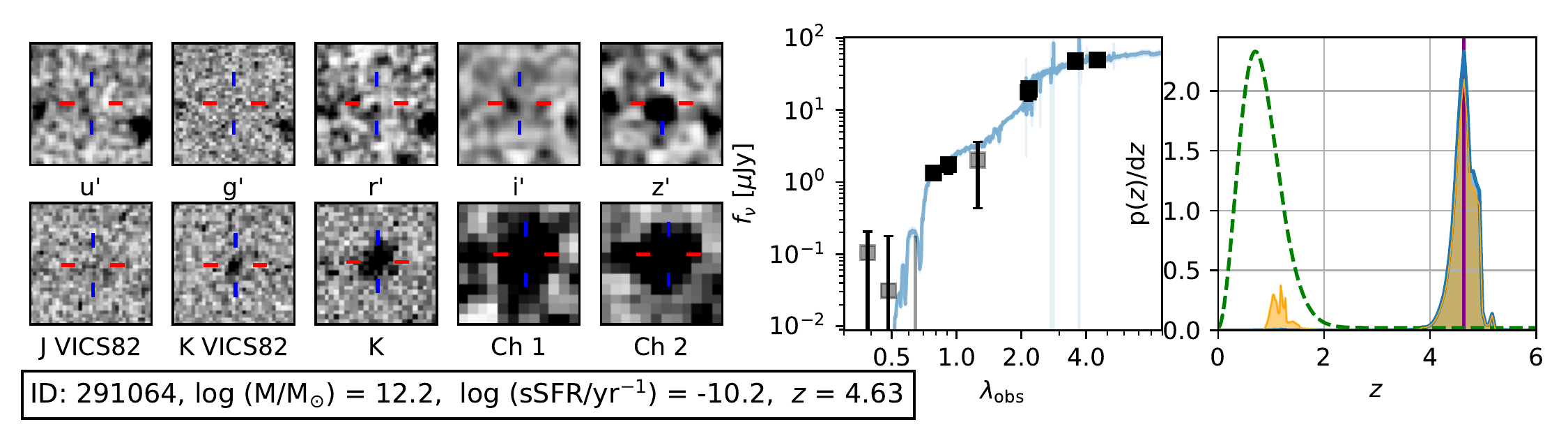}
\includegraphics[scale=0.6,angle=0]{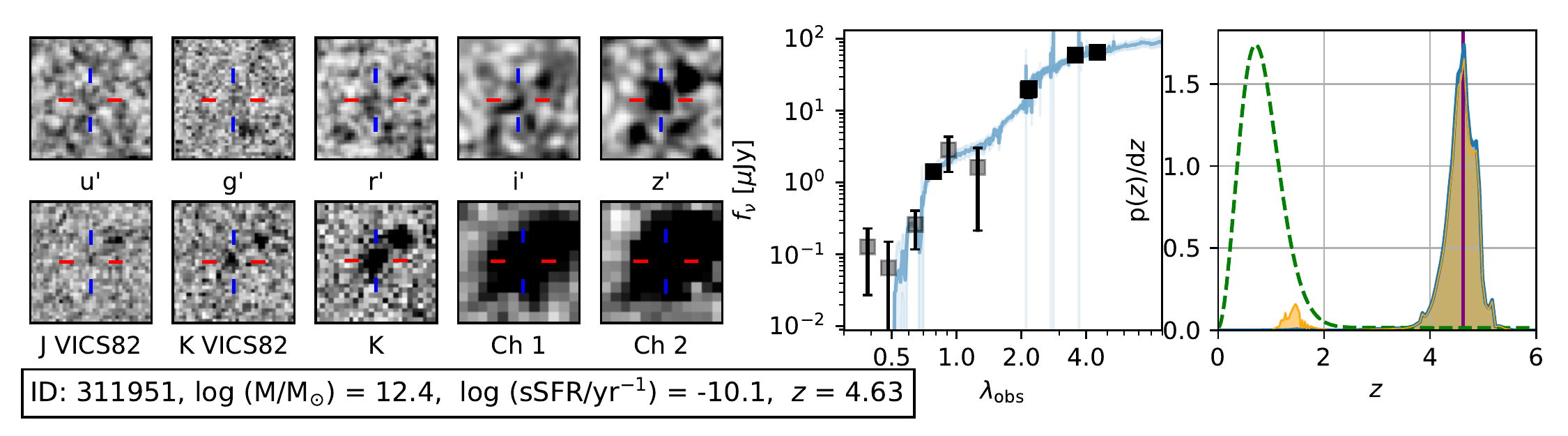}
\includegraphics[scale=0.6,angle=0]{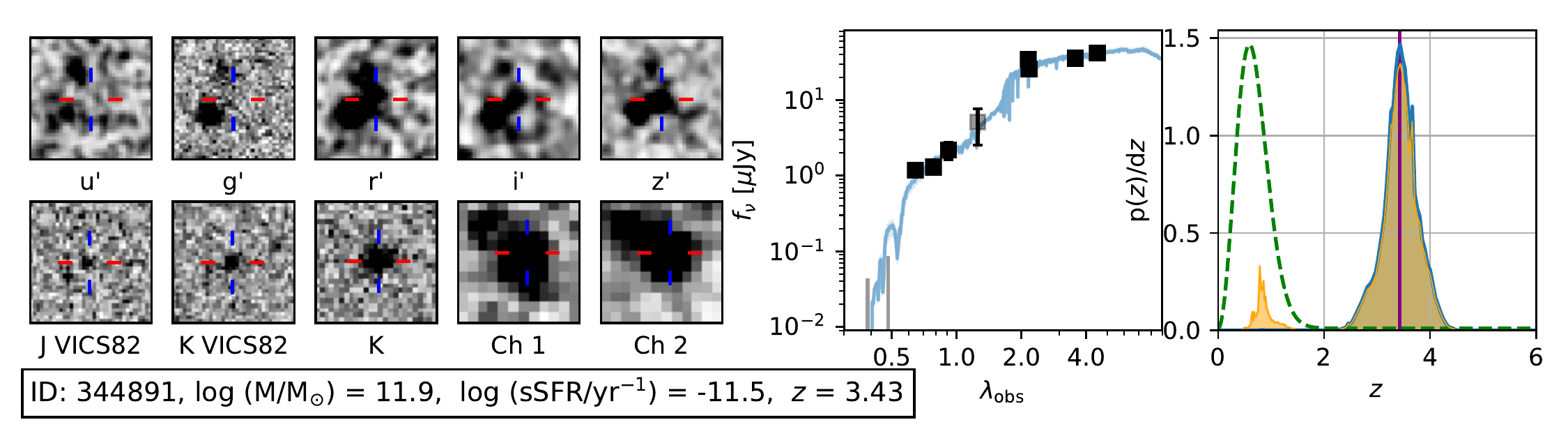}
\includegraphics[scale=0.6,angle=0]{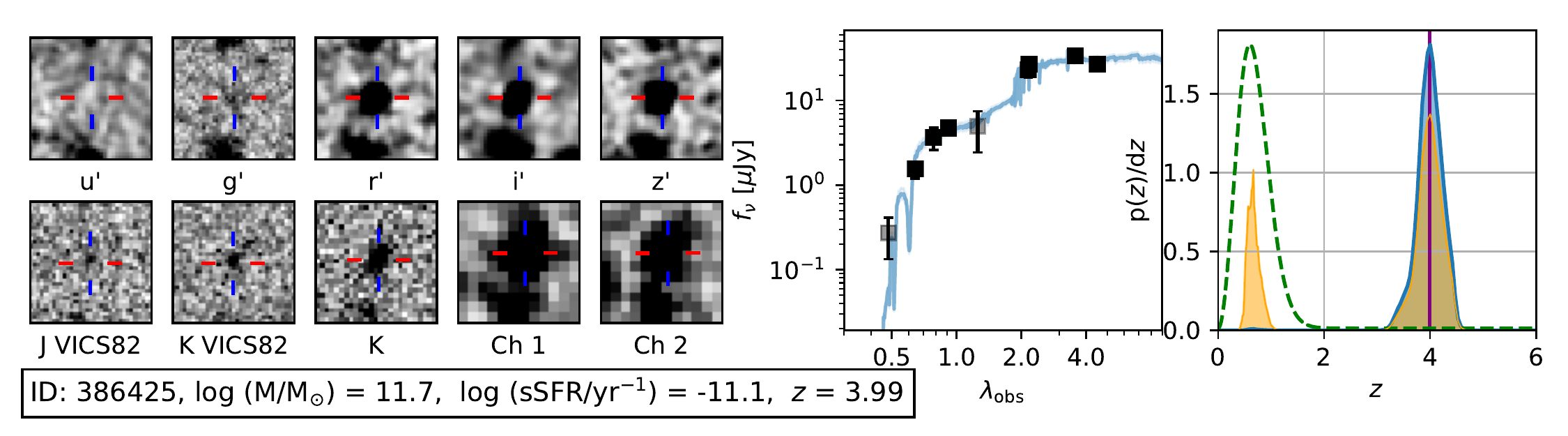}
\caption{For the first five of nine of our massive, quiescent, $z=3-5$ galaxy candidates: photometric images (left), best-fitting templates with the photometry (center), and $z$PDF (right). The images are 10$''$x10$''$ in all bands. The squares in the middle plot show the observed photometry, shown in gray when the S/N is less than 2. \edit2{The blue-shaded curve shows the measured P($z$) \emph{before} convolving with the prior (shown in green), while orange shows the result after convolution, with the vertical line denoting the best-fitting redshift after convolution.}}
\label{Fig:stamps-q}
\end{figure*}

\begin{figure*}[!t]
\centering
\includegraphics[scale=0.6,angle=0]{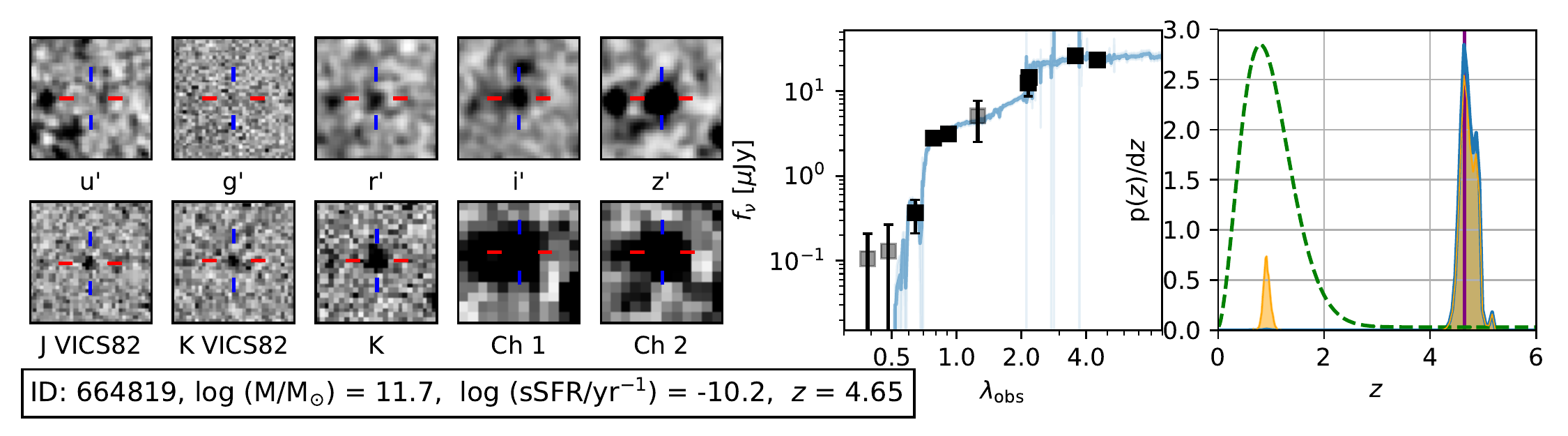}
\includegraphics[scale=0.6,angle=0]{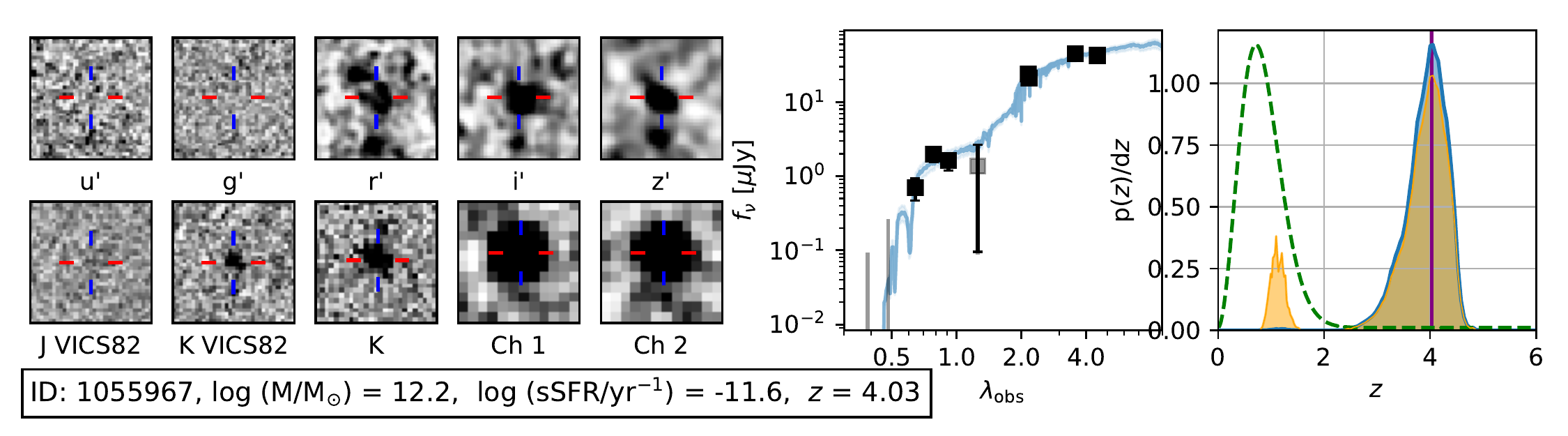}
\includegraphics[scale=0.6,angle=0]{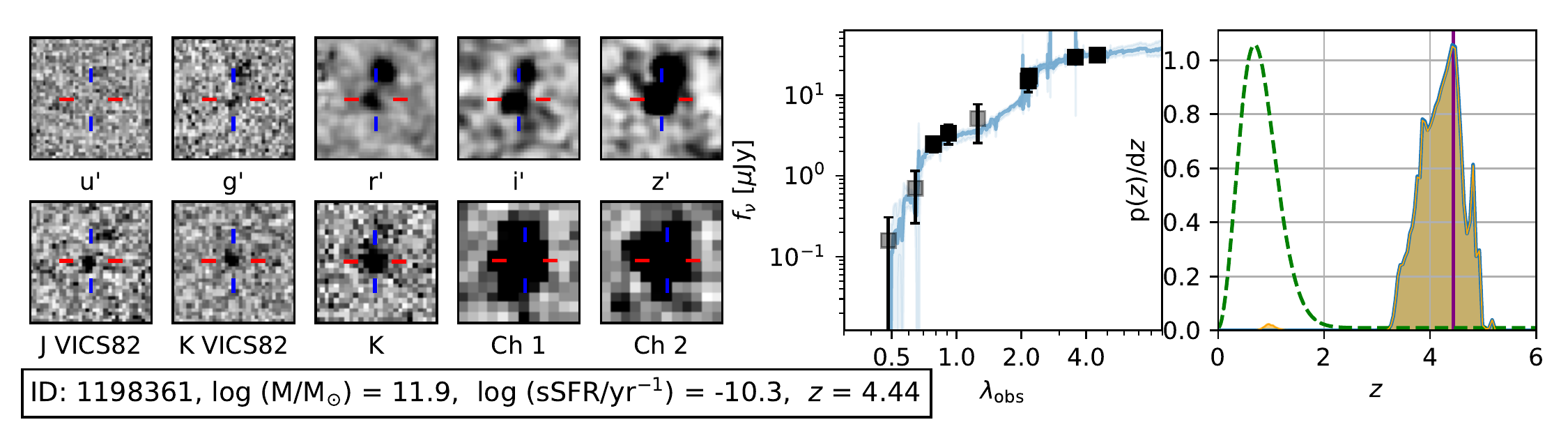}
\includegraphics[scale=0.6,angle=0]{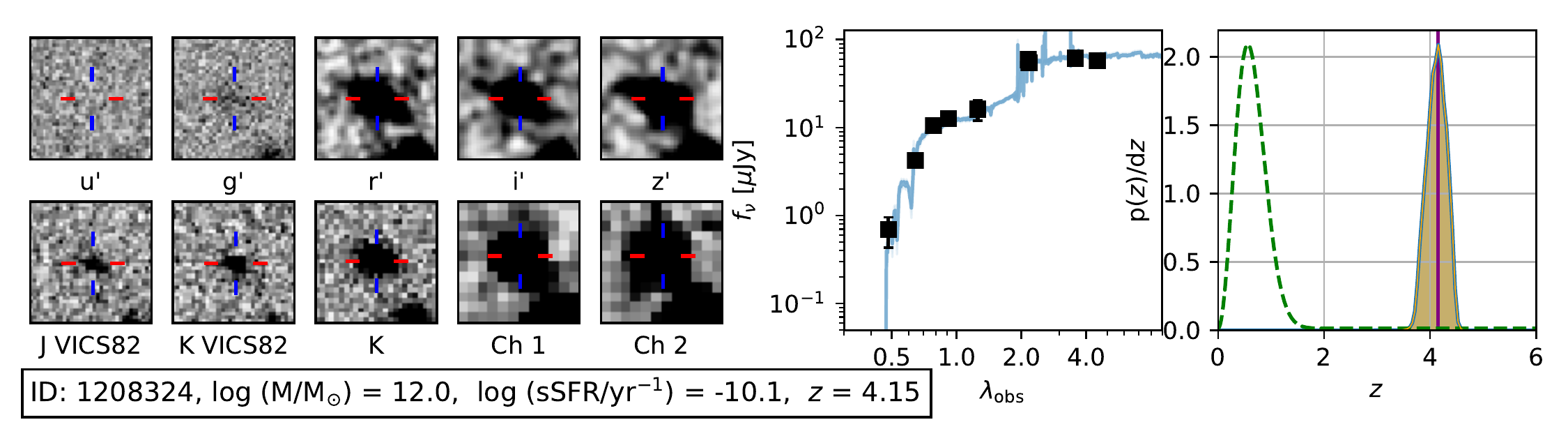}
\caption{Same as Figure~\ref{Fig:stamps-q}, for the final four candidates.}
\label{Fig:stamps-q2}
\end{figure*}

While we chose to use EAZY-py to fit the stellar population properties and photometric redshifts for its speed (as we ran this on our full catalog), EAZY-py does not calculate uncertainties on the physical properties.  This is not desirable, as some properties, particularly sSFR, may have wide posteriors.  We thus ran a second iteration of SED fitting on these 51 candidate massive quiescent galaxies, both to vet the results from EAZY-py, and also to generate parameter posterior distributions.  For this second method, we used the grid-based Monte Carlo SED fitting method developed by \citet{finkelstein15}, utilizing the stellar population models of \citet{bc03}.  This method finds the best-fit via a maximal likelihood search of the parameter space, deriving posterior distributions on those parameters via a Monte Carlo analysis.  The addition of this analysis allows us to test how well our derived stellar population properties agree when using both different stellar population models, and a different fitting method.

The agreement in stellar mass between these two methods is very good, with all 51 candidates having log(M/M\sol) $>$11, with the mass difference $<$0.3 dex for 90\% of the sample, with a maximum difference of 0.6 dex.  The difference in SFR is however larger, with EAZY-py resulting in systematically higher SFRs by $\sim$0.5 dex, and only $\sim$30\% of the sample having SFR differences of $<$0.5 dex.  Due to the potential for large differences in the inferred SFR from the use of different modeling codes and templates, we further hone our candidate list by requiring galaxies to have log(M/M\sol) $>$ 11 and log (sSFR/yr$^{-1}$) $< -$10.05 from both methods.  This reduces the sample from 50 to 23 candidates.  For this reduced sample, the SFRs agree very well with a median difference of zero, and a standard deviation of 0.6 dex.

To further examine the validity of these candidates, we make use of the posterior distributions made available by the additional BC03-based SED fitting to explore how confident we can be that these sources are quiescent.  We calculated the integral of the sSFR posteriors which were below our sSFR cut.  We found that most sources had relatively high values, indicating tight sSFR posteriors centered at low values, but there were a few with very wide sSFR posteriors leading to small values of this integral.  We thus apply one final cut to our sample, requiring that $>$68\% of the sSFR posterior satisfy our sSFR cut ($<-$10.05).  This removes five additional sources, leaving 18 candidate quiescent galaxies.

As a final quality check on our candidate galaxies, we visually inspected these 18 candidates.  We found that five candidates had potential problems with their photometry - four had visible smudges in the $u$-band, which while not formally exceeding 2$\sigma$, visually appeared above the noise and coincident with the source position, while one source had no VISTA measurements due to bad pixels in both $J$ and $K$ at the source position, leaving the 4000\AA\ break poorly constrained. Four more sources appear visibly elongated, and have half-light radii (r$_h$) measurements $\gtrsim$1$^{\prime\prime}$, and are thus significantly resolved (bright, unresolved sources in this catalog have r$_h =$ 0.7 $\pm$ 0.1$^{\prime\prime}$).  This corresponds to a physical half-light radius of $\sim$7 kpc at these redshifts, highly unlikely for massive galaxies at $z \sim$ 4 as trends at lower redshifts would predict r$_h$ $<$ 1 kpc \citep[e.g.,][]{vanderwel14}.  While these nine sources formally satisfy our selection criteria, this visual inspection casts doubt on their validity, thus we remove them from our high-confidence sample.  All nine of these removed candidates are shown in the Appendix in Figures~\ref{Fig:stamps-q-appen1} and \ref{Fig:stamps-q-appen2}, and are listed in Table~\ref{qgcat-appen}.

We show our final sample of nine candidate massive quiescent galaxies in Figures~\ref{Fig:stamps-q} and \ref{Fig:stamps-q2}, with relevant quantities for these galaxies listed in Table~\ref{qgcat}.

\subsubsection{Comparison to UVJ Selection}

\begin{figure}[!t]
\centering
\includegraphics[scale=0.6,angle=0]{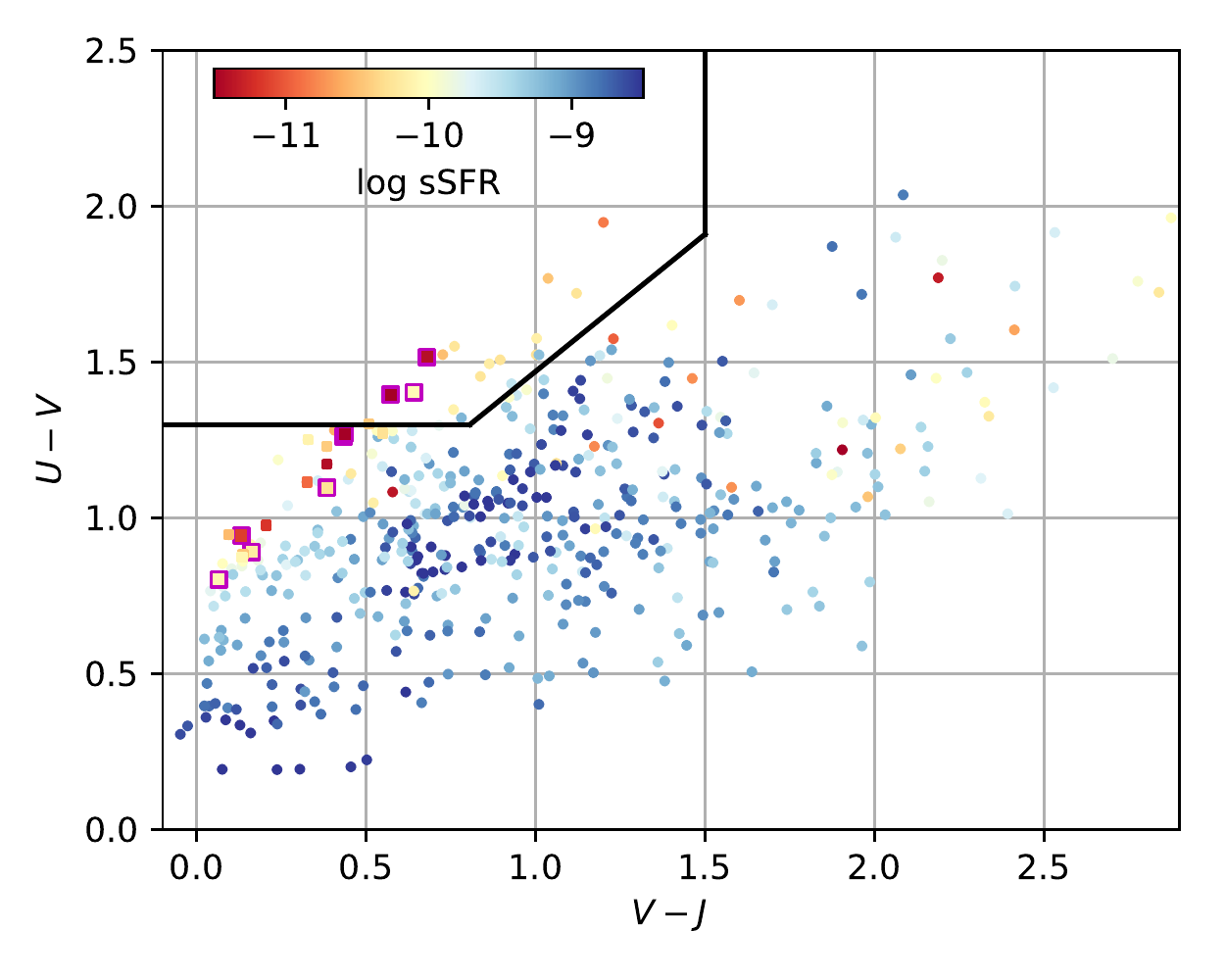}
\caption{$UVJ$ diagram for massive (log M/M\sol\ $>$ 11) at 3 $\leq z <$ 5. The symbol color encodes the sSFR from EAZY-py.  Squares denote our massive quiescent candidates (\edit2{with those outlined in larger magenta boxes} coming from the final higher-confidence sample of nine sources), while circles denote star-forming galaxies.   A substantial portion of the quiescent galaxies selected via a sSFR threshold fall outside of the selection box from \citet{muzzin13}, suggesting a custom selection box would be required for our sample. However, such a selection box would also include a significant number of star-forming galaxies, highlighting the utility of our sSFR-based selection.}
\label{Fig:uvj}
\end{figure}

In Figure \ref{Fig:uvj}, we show our $z=3-5$ massive quiescent galaxy candidates in two $U-V$ vs $V-J$ diagrams to inspect where they fall with respect to a relevant $UVJ$ selection box. We show both the nine higher-confidence sources, the nine sources we removed to the appendix, and the full parent sample of 466 massive galaxies.  The color of the symbols denote the log sSFR/yr$^{-1}$. In each panel we include the selection box from \citet{muzzin13} for reference. The quiescent galaxies form a locus that is offset from the diagonal part of the reference selection box (bluer in both colors by up to 0.5 mag) with the majority of our sSFR-selected quiescent galaxies falling outside of the $UVJ$ selection box. We also see that a fraction of objects with log sSFR/yr$^{-1}>-11$ fall into the $UVJ$ selection box, so had we used this $UVJ$ selection box to select quiescent galaxies our SED modeling implies that we would have been very incomplete to quiescent galaxies and selected a number of star-forming galaxies.

This issue with the $UVJ$ diagram has been seen before. A study of passive galaxies at $z > 3$ in the GOODS-South field by \citet{merlin18}, who selected passive galaxies using SED fitting, found an offset between their sSFR-selected passive galaxy candidates and the nominal $UVJ$ diagram selection box. They note that a $UVJ$ selection of passive galaxies is incomplete, and that if a top-hat SFH (where galaxies form stars continuously for some time, and then quench) is assumed, galaxies are quiescent for $\sim$0.5 Gyr before they enter the $UVJ$ selection box from below, similar to the locus of our candidate sample. This can be seen in Fig \ref{Fig:uvj}, where the candidates with the lowest sSFR fall in the selection box, but those closer to our log sSFR/yr$^{-1}>-11$ threshold (which likely have had less time since quenching than the lowest sSFR objects) fall outside of it.  Since galaxies at $z >$ 4 exist when the Universe is $<$2 Gyr old, it is not surprising that these quiescent galaxies may not have been quenched for $>$0.5 Gyr.  In this sense, our galaxies may be more analagous to post-starburst galaxies local rather than fully quenched galaxies.  \edit2{Similar conclusions have been reached by other studies in this epoch \citep[e.g.,][]{marsan20,deugenio20,forrest20b}.}

\subsection{Properties of Candidate Massive Quiescent Galaxies} \label{subsec:props}

As listed in Table~\ref{qgcat}, and shown in the right panel of each candidate's plot, the P($z$) distributions for our nine candidates strongly prefer high-redshift.  Though we required $\int$P($z>$3) $>$ 0.6, 7/9 candidates have this quantity $>$0.8.  This is understandable due to our use of the prior, shown as the green line in the P($z$) panel for each object.  Objects which have even small P($z$) secondary peaks where this prior is significantly non-zero will have P($z$) which strongly prefers a low-redshift solution after folding in the prior, hence these objects do not satisfy the sample-selection criteria.

To enable both robust photometeric selection, and decent constraints that a galaxy is quiescent, we would expect the SED to show a Lyman break (e.g., no significant flux in the $u$-band, and a decrement in the $g$-band), weak non-ionizing UV flux, and a significant Balmer break, with a roughly flat SED redward of 2 $\mu$m.  Inspecting the candidates, this description is broadly characteristic of our sample.  The best-fit model follows the photometry closely, which is expected given our stringent $\chi^2$ cut.

By definition, our sample should have large stellar masses, and given our criteria selecting for highly robust quiescent solutions over a very large volume, we should expect to be discovering amongst the most massive galaxies in this epoch.  However, it is surprising that we find several (5/9) galaxies with log M/M\sol\ $>$ 12.  Upon investigation of the stamp images, several of the sources do show nearby neighbors, which could be affecting the photometry and subsequent stellar mass values.  We found that every neighbor was present in the input catalog, either due to being detected in the NHS $K_s$ catalog, or in the DECam (for very blue sources) or IRAC (for very red sources).  This ensures that TPHOT is modeling each source separately.  We are confident that the ground-based fluxes are accurate, as these neighbors appear well resolved.  This is less certain in IRAC due to the larger PSF.  While TPHOT is designed for these situations, it is possible that the IRAC measurements of our sources contain some residual flux from these neighbors, biasing the stellar masses high.  Even if the deblending is accurate, it is possible that these neighbors are gravitationally lensing our candidates by a factor of $\sim$a few \citep[e.g][]{mason15} due to their close proximity.  Finally, galaxies this massive may have some of their light contributed due to AGN activity.  Though this is unlikely to dominate the luminosity due to the stellar appearance of the SED, it could also contribute primarily to the IRAC fluxes, biasing the masses high.

Although we have discussed a few reasons why these extremely massive galaxies may have true stellar masses lower by a factor of $\sim$a few, it is worth exploring how surprising their presence is.  Ideally we would compare to published stellar mass functions from observations or simulations, but as we are the first to explore galaxies in this epoch over such a large volume, such an idea comparison does not exist.  Perhaps the easiest comparison is to \citet{sherman20}, who used this same NHS catalog to select galaxies up to $z \sim$ 3.  Over our same area, they found 198 and 12 galaxies at log(M/M\sol) $>$ 11.5 and 12.0, respectively, at $z =$ 3.25.  \citet{muzzin13} calculated the stellar mass function at $z =$ 3.5 over 1.6 deg$^2$, with these functions predicting 692 and 22 galaxies at log(M/M\sol) $>$ 11.5 and 12.0, respectively, over our area of 17.5 deg$^2$.  Finally, \citet{davidzon17} calculated the stellar mass function at $z =$ 4 over 2 deg$^2$, which would predict 59 and 0 galaxies at log(M/M\sol) $>$ 11.5 and 12.0, respectively, over our area.  

Collectively, these previous results show that the presence of $z >$ 3 galaxies at log(M/M\sol) $>$ 12.0 is not unprecedented, but the large dispersion in these results suggests that their abundance is not well quantified. We also note that these predicted numbers rely on the extrapolation of a Schechter function, which assumes an exponential cutoff, while the shape of the massive end of the stellar mass function hasn't been observationally validated at these redshifts. On the theory side, \citet{behroozi18} found that in a SHELA-like survey at $z =$ 4 the presence of several galaxies with log(M/M\sol) $>$ 12 does not violate constraints on the growth of structure from $\Lambda$CDM (extrapolating their Figure 2 to $z =$ 4; P.\ Behroozi, private communication).  Additionally, the {\sc Simba} numerical hydrodynamics simulation \citep{dave19} contained 3 galaxies with log(M/M\sol) $>$ 12 at $z =$ 3 over a relatively small volume (143$^3$ cMpc$^3$).  None this massive were present at $z =$ 4, though the simulation did contain five $z =$ 4 galaxies with log(M/M\sol) $>$ 11.5.  

While observations and theory thus don't prohibit the presence of these massive galaxies, clearly further observations are needed to better prove the validity of these sources.  
First and foremost, confirmation via spectroscopy that these sources are at the estimated redshifts would remove the strongest concern, that the sample is contaminated by lower-redshift sources.  As these galaxies still have some low-level residual star formation, this may be possible via [O\,{\sc ii}] emission with ground-based $K$-band spectroscopy.  Lacking emission lines, confirmation via stellar absorption lines is possible, though this will likely require the sensitivity of the {\it James Webb Space Telescope} ({\it JWST}).  Further improvement on the SED constraints are also possible, as better removal of low-redshift solutions (needed to keep true candidates from being removed with application of the magnitude prior) can be obtained with deeper $u$ and $g$-band imaging.  Likewise, better constraints on the star-forming properties can be obtained with deeper near-infrared imaging with, for example, FLAMINGOS-2 or NIRI on Gemini in the near-term, or {\it JWST} in the longer-term.  Finally, dust-continuum observations with ALMA can rule out both low-redshift dusty contaminants, as well as significant obscured star-formation in $z \sim$ 4 galaxies.  The combination of the large volume probed by our catalog with these followup observations will ultimately allow large samples of confirmed massive quiescent galaxies to be formed, challenging theoretical models of galaxy formation.

\subsection{Comparison with Previous Results} \label{subsec:compresults}

At $z \sim$ 2--3, using this same NEWFIRM catalog \citet{sherman20b} and \citet{florez20} explored the quiescent fraction amongst massive galaxies, and found it to be $\sim$20-25\% using a similar selection as we employed.  Moving to $z >$ 3, \citet{muzzin13} performed the previous widest-area search for quiescent galaxies (1.6 deg$^2$), using a UVJ selection to find 28 massive quiescent galaxies and 146 star-forming galaxies with log M/M$_{\odot} >$ 10.94 at $3\leq z<4$.
\citet{stefanon13} found four candidate massive compact quiescent galaxies at $z >$ 3, though their stellar mass threshold was lower due to their smaller volume (the search encompassed the CANDELS fields).  \citet{spitler14} used UVJ selection to find that roughly \emph{half} of a sample of log (M/M\sol) $>$ 10.6 galaxies were quiescent using the zFourGE survey over a subset of the CANDELS fields (see also \citealt{straatman14}).  

Studies have also recently been successful spectroscopically confirming these massive quiescent galaxies at $z \sim$ 3--4 and ruling out significant levels of obscured star-formation with ALMA \citep[e.g.][]{glazebrook17,valentino20,santini19,marsan20,schreiber18,tanaka19,deugenio20,forrest20a,forrest20b}.  Though these surveys have different selection techniques, mass limits, and volumes, together they collectively show that massive quiescent galaxies exist in some non-zero abundance at $z >$ 3.

We compare our measurements directly to predictions from two recent simulations.  The first is the updated ``Santa Cruz'' semi-analytic model \citep[SC-SAM][Somerville et al., in prep]{somerville15,somerville08}. The SC-SAM includes standard recipes for the main processes in galaxy formation, such as gas cooling, star formation, and stellar feedback, and implements AGN feedback in two modes: thermal energy deposition from a low-accretion rate, radiatively inefficient ``jet mode'' (which can offset cooling flows), and ejection of cold interstellar gas via radiation pressure from rapidly accreting, radiatively efficient black holes \citep[see][for more details]{somerville08}. To calculate the quiescent fraction from this model, we made use of the lightcones created for the CANDELS project, which consist of eight realizations of each of the five CANDELS fields, for 40 realizations total.  The combination of these lightcones results in a total area of $11.9$ square degrees, corresponding to a substantial volume of 0.28 Gpc$^{3}$ at 3 $< z <$ 5, or roughly 70\% of what we probe observationally\footnote{However, it should be kept in mind that these lightcones were drawn from an N-body simulation with a total volume of $4.56\times 10^7$ Mpc$^3$, and unlike in the real Universe, the structures along the line of sight are not independent.}. This yields 576 SAM galaxies in our redshift and mass range of interest.  As all of the SAM galaxies have log M/M$_{\odot}$ $<$ 11.4, we can only compare over a narrow mass range.  Of the 576 SAM galaxies, only 25 meet our definition of quiescence, for a quiescent fraction of 4.3\% from this simulation.  However, all of these quiescent sources have $z <$ 4, thus this model is in tension with our observations in that it both produces no quiescent galaxies at $z >$ 4.

We also compare to recent results from the {\sc Simba} numerical hydrodynamics simulation \citep{dave19}.  {\sc Simba} is the successor to the {\sc Mufasa} simulations, and is run with the GIZMO meshless finite mass hydrodynamics scheme. {\sc Simba} included updated sub-grid prescriptions for star formation and stellar driven outflows, and one of the most sophisticated and physically motivated treatments of black hole growth and AGN feedback in any available cosmological simulation. Like the SC-SAM, {\sc Simba} implements two modes of AGN feedback, which are differentiated based on the accretion rate of the black hole. An important caveat is that the volume of this simulation is rather limited, $\sim (143 Mpc)^3 \sim 3 \times 10^{6}$~Mpc$^{3}$, which is $\sim$1\% of our survey volume.  At $z=3$, {\sc Simba} has 109 massive galaxies with log M/M$_{\odot}$ $>$ 11.  However, only 10 of these galaxies have sSFR $<$ 10$^{-10}$ yr$^{-1}$.  At $z =$ 4, there are 30 massive galaxies, though none meet our definition of quiescence, while there are still a significant number of observational counterparts. We discuss possible interpretations of the discrepancy between both theoretical predictions and the observational results in Section~\ref{sec:discussion}.

\section{Discussion} \label{sec:discussion}

Our discovery of a \edit1{potential} population of massive quiescent galaxies at $z=3-5$ has significant implications for understanding the physical processes that shape the formation and evolution of galaxies.  \edit1{If followup observations confirm} the existence of massive quiescent galaxies at $z\sim 4$--5, when the universe was only 1.2-1.5 Gyr old, implies that the universe can rapidly build galaxies 1-10 times the mass of the Milky Way and quench their star formation in a very short time. If the first galaxies start to form around 0.3-0.4 Gyr after the Big Bang, this leaves less than 1 billion years for massive quiescent galaxies at $z\sim 5$ to grow. It is unclear whether the same mechanisms that are responsible for forming quiescent galaxies at $z=0-2$ can be responsible for forming their counterparts at $z>5$.

Several previous studies have examined theoretical predictions for
quiescent fractions at $z=0-2$ and identified a number of mechanisms
or formation paths that may be responsible. \citet{kimm09} analyzed several different semi-analytic models and quantified the fraction of quiescent galaxies as a function of internal properties and environment, compared with group catalogs extracted from the Sloan Digital Sky Survey. 

\citet{brennan15} compared
observations of the evolution of the massive quiescent galaxy fraction
(using their own analysis of the CANDELS fields, finding a
quiescent fraction of $\sim$30\% at $z =$ 3) to a previous generation
of the Santa Cruz semi-analytic model, where passive spheroids are
built up via AGN feedback originating from major and minor mergers and
disk instabilities.  In both observations and simulations, rather than
a constant threshold in sSFR, they imposed a cutoff based on the
offset in SFR from the star-forming main sequence at a given
redshift.  Their simulations predict a quiescent fraction which is in
reasonable agreement with $z \sim$ 0 observations, but drops to zero
by $z \sim$ 2.5, becoming increasingly discrepant with observations at
increasing redshifts.

In the hydrodynamical cosmological simulation Illustris, \citet{wellons15} investigated the formation of compact massive galaxies in a box of 106.5 co-moving Mpc on a side. They found two main formation paths for compact massive galaxies at $z=2$: (1) early formation when the universe was denser, and (2) a central burst of star formation from a major merger. The first path consisted of formation early in the universe (log M/M$_{\odot}$ $\sim$ 10 before $z\sim7$) and smooth stellar mass growth at a rate less than the typical galaxy of similar mass until $z\sim4$ when the stellar mass exceeded log M/M$_{\odot}$ $=$ 11. At that time, the feedback from the SMBH began to significantly decrease the SFR, quenching the galaxy by about $z\sim2.5$. In the simulation, this path produced a massive galaxy at $z\sim4$, however, the galaxy was not quenched (log sSFR $< -11$ yr$^{-1}$) until $z\sim2.5$ after 1 Gyr of SMBH feedback. The second path involved an intermediate mass galaxy growing at a typical SFR for its mass, from early times to $z\sim2.5$ when a wet major merger caused the SFR to spike and the stellar mass to double to log M/M$_{\odot}$ $>$ 11. The wet major merger also caused accretion onto and feedback from the central SMBH decreasing the SFR. However, once again the decrease is gradual; if we were to apply these scenarios to our observational sample, we would indeed be able to create massive galaxies by the $z \sim$ 5 epoch, but the objects would not be classified as quiescent until well after our redshift window.

The successor of the Illustris simulations, IllustrisTNG, included an updated kinetic implementation of ''jet mode'' AGN feedback \citep{weinberger17,weinberger18}, in part to correct the weak color distribution bimodality between red and blue galaxies found in the original Illustris simulations \citep{nelson18}. \citet{tacchella19} explores the build-up of spheroids and discs in IllustrisTNG and finds the most massive (log M/M$_{\odot}$ $=$ 11.5) galaxies at $z=0$ grow more slowly in situ, not reaching log M/M$_{\odot}$ $>$ 11 until $z\sim1$. In these galaxies, SMBH feedback becomes effective at ceasing star formation when the SMBH mass exceeds log (M$_{BH}$/M$_{\odot}$) $\sim$ 8.2 around $z\sim2$, when the growth of stellar mass becomes dominated by mergers. The faster and more efficient SMBH feedback in IllustrisTNG can quench massive galaxies quickly, however, the implementation of strong feedback at all times appears to slow the in situ growth of galaxies and prevent massive systems from forming at early times. Figure~7 of \citet{merlin19} shows the number density of quiescent galaxies with stellar mass greater than $5 \times 10^9$ M$_{\odot}$ extracted from the original Illustris simulations and from the IllustrisTNG 100 and 300 boxes, defining quiescence as sSFR $<$ 10$^{-11}$ yr$^{-1}$. With the major caveat that the stellar mass limit of the sample they have extracted from simulations is \emph{much} lower than the one we have been considering, they show that IllustrisTNG has a significantly larger number density of quiescent galaxies at $3 \lesssim z \lesssim 4$ than the original Illustris or {\sc Simba} simulations, but the number of quiescent galaxies drops steeply at $z>3.5$. Their comparison shows that the EAGLE simulations \citep{Schaye:2014} predict the highest number density of quiescent galaxies at $z\gtrsim 3.5$ of any of the simulations they considered.

Non-black-hole-related mechanisms have also been proposed for quenching massive galaxies, including morphological quenching and environmental quenching.  Morphological quenching is the suppression of star-formation in a gaseous disk inside a spheroidal galaxy due, in part, to the tidal forces of the stars preventing fragmentation in the gaseous disk.
\citet{martig09} demonstrated this mechanism with a massive galaxy growing relatively slowly at low redshift, from log M/M$_{\odot}$ $=$ 10.5 at $z=2$ to log M/M$_{\odot}$ $=$ 11 by $z\sim0.8$, spanning $\sim$4 Gyr. For morphological quenching to work in the early universe, it would have to work on much shorter time scales than has been demonstrated, though high-resolution imaging of our sample could at least determine if they even have spheroidal morphologies.
Environmental quenching via ram pressure or tidal stripping of gas can operate on satellite galaxies in dense environments, but the majority of galaxies in our sample are unlikely to fit these criteria.

There is now a general consensus that physical processes related to the radiation and kinetic energy emitted by accreting SMBHs (AGN feedback) are the dominant mechanism for quenching star formation in massive galaxies. For example, \citep[e.g.][]{su19} showed that previously proposed processes such as gravitational heating and morphological quenching are not viable on their own --- only SMBH-related processes can lead to long-lived quenching -- reducing the SFR well below the star-formation main sequence and keeping it there.

Most current semi-analytic models, as well as most numerical cosmological hydrodynamic simulations (e.g., IllustrisTNG, HorizonAGN, Romulus) make assumptions that cause feedback from radiatively efficient accretion onto a SMBH to be inefficient at quenching, or to quench only over a short timescale. Feedback from a low-accretion rate, radiatively inefficient ''jet mode" is responsible for strong, sustained quenching in massive central galaxies in both the Santa Cruz SAM and in IllustrisTNG. However, more detailed modeling of ``radiative" feedback from more rapidly accreting black holes has shown that this process may drive powerful, high velocity winds (as observed ubiquitously in AGN hosts) which can not only remove the cold dense gas from the interstellar medium of galaxies, leading to strong rapid quenching, but also prevent further accretion of gas over timescales of many Gyr \citep{choi15,choi17,brennan18,pandya17}. To date these simulations have only been carried out as ``zoom-ins"  of a few tens of massive halos, so it is not possible to compute a statistical quenched fraction. However, it is clear that this feedback mode will be effective at higher redshifts than the ``jet mode," which requires more massive black holes, lower accretion rates, and the development of a hot circumgalactic halo (``working surface"), all of which develop only at later cosmic epochs.
Some hydrodynamic simulations, such as EAGLE and {\sc Simba}, have included an implementation of the ``strong radiative mode" quenching in cosmological volumes.
{\sc Simba} produces a population of ``fast-quenched" galaxies which are more prevalent at early epochs \citep{2019MNRAS.490.2139R}, but the early quenching is typically not sufficient to drop below sSFR=10$^{-10}$ yr$^{-1}$ at $z >$ 4, even though it drops the sSFR by an order of magnitude relative to the main sequence.
Currently these simulation volumes tend to be fairly small, ($\sim 100$ Mpc on a side), so it is not yet possible to robustly study the statistical properties of such massive galaxies as the ones in this study. 

It is important to note when comparing results that different observational and theoretical studies often employ different definitions of quiescence.  Most deal with the degree of quiescence, either as we do by imposing a single sSFR threshold, or as \citet{brennan15} have done by looking at how reduced the SFR is compared to the main sequence, which takes into account the redshift dependence of the SF main sequence normalization.  Another important question is the timescale of this quiescence -- how long did it take for the galaxies to become quiescent --- is the process of quenching slow or fast? Are the galaxies permanently quenched, or will they become star-forming again in the future?  The former can be addressed empirically by constraining the star-formation histories of our quenched galaxies to explore how long they have been quenched, which we will address in future work. Constraints on quenching timescales can provide even more powerful tests of the physical processes responsible for quenching in theoretical models \citep{pandya17}.

\section{Summary} \label{sec:summary}
In this paper, we have presented the $K_{s}$-selected catalog of the NEWFIRM HETDEX Survey (NHS) within the SHELA field. The catalog has a moderate depth ($K_{s} \sim$ 22.\edit1{4} mag, 5$\sigma$) and is unique in covering a wide $\sim$22 deg$^2$ area. When combined with the existing extensive multi-wavelength dataset in the SHELA field, the \edit2{volume probed by the} NHS data allows for the detection of the most statistically significant population of massive galaxies at high redshift assembled to date.

We derive photometric redshifts, stellar masses and rest-frame colors for massive (log M/M$_{\odot}$ $>$ 11), star-forming and quiescent galaxies between $z=3-5$ using the SED-fitting code \pkg{eazy-py}. We select galaxies using a procedure utilizing the zPDF and define quiescence with a sSFR threshold (log sSFR/yr$^{-1}<$ -10.05; 3$\sigma$ below the $z =$ 4 main sequence of \citealt{salmon15}). \edit1{We utilize an extremely conservative (allowing for potential incompleteness) selection process to identify a sample of nine massive quiescent candidates out to $z \sim$ 4.7.}  We locate our sSFR-threshold-based quiescent galaxy sample on the $UVJ$ diagram and find our sample occupies a parameter space not typically selected by past studies utilizing the UVJ selection method, \edit1{suggesting these candidate galaxies are post-starburst in nature, and not yet fully quenched.}
These results imply that there exists a population of massive galaxies at even higher redshifts that have rapidly formed their stars and have been quenched in less than $\sim$1.5 Gyr since the Big Bang.

We compare our results with the predictions of two different theoretical cosmological simulations of galaxy formation, which incorporate physical prescriptions for feedback from accreting SMBHs: the Santa Cruz semi-analytic model and the {\sc Simba} hydrodynamic simulation. Both simulations reproduce the low-redshift population of massive, quiescent, spheroid-dominated galaxies, but significantly underpredict the presence of massive quiescent galaxies at $z >$ 4. We speculate that either quenching is not effective enough, begins too late, or operates too slowly in these theoretical models. 

The results presented in this work thus strongly suggest that our understanding of the physical processes that quench massive galaxies at high redshift is incomplete, and requires further investigation.  To have full confidence in this result we must overcome the impact of contamination from dusty low-$z$ interloping galaxies.  Observations from ALMA and deep IR imaging and spectroscopy from the \textit{James Webb Space Telescope} (\textit{JWST}) are needed to probe for obscured star formation, and spectroscopically identify the strong 4000 \AA\ break in quiescent galaxies, respectively.

\acknowledgments
M.L.S.\ and S.L.F.\ acknowledge support from the National Science Foundation through grant AST 1614798 and from the NASA Astrophysics and Data Analysis Program through grants NNX16AN46G and 80NSSC18K0954. S.S.\, S.J.\, and J.F.\ gratefully acknowledge support from the University of Texas at Austin, as well as NSF grants  AST-1413652, AST-1614798, and  AST-1757983 .  The work of CP and LW is supported by the NSF grant AST 1614668.  J.F.\ acknowledges support from NSF GRFP through the grant DGE-1610403. RSS and LY acknowledge support from the Simons Foundation and the Downsbrough family. We thank the anonymous referee for their constructive comments.  We thank Paola Santini, Rachel Bezanson, Justin Spilker, Eric Gawiser, Raquel Martinez, Neal Evans, and Gabe Brammer for useful conversations which improved the quality of this work. We thank Mark Dickinson, David Herrera, Ron Probst, Frank Valdes, and all of the NOAO scientists and staff who helped make this survey possible. We thank Yi-Kuan Chiang, Yaswant Devarakonda, Emily McLinden, Mallory Molina, Sofía Rojas, Heath Shipley, Mimi Song, Vithal Tilvi, Rebecca Tippens, Tim Weinzirl, and Greg Zeimann for volunteering their time to observe with NEWFIRM onsite at Kitt Peak National Observatory.

This work was based on observations at Kitt Peak National Observatory, NSF's National Optical-Infrared Astronomy Research Laboratory (Prop. ID 13B-0236; PI: S. Finkelstein), which is operated by the Association of Universities for Research in Astronomy (AURA) under cooperative agreement with the National Science Foundation. The authors are honored to be permitted to conduct astronomical research on Iolkam Du'ag (Kitt Peak), a mountain with particular significance to the Tohono O'odham.

PyRAF is a product of the Space Telescope Science Institute, which is operated by AURA for NASA.

\software{Source Extractor (Bertin \& Arnouts 1996), PSFEx (Bertin 2011), eazy-py (https://github.com/gbrammer/eazy-py), NEWFIRM calibration pipeline (Swaters et al. 2009), pythonFSPS (Foreman-Mackey et al. 2014), Tractor (Lang et al. 2016a,b)}

\facilities{Mayall (NEWFIRM), Blanco (DECam), Spitzer (IRAC)}

\bibliographystyle{aasjournal}

\appendix

\edit3{In this appendix, we first list the estimated 50\% and 90\% $K$-band completeness for each of the NHS tiles.  This was estimated by inputting 2000 mock point-sources in each tile, re-running our photometric measurement process in the same way as in our actual catalog, and calculating the fraction of input sources which were recovered in magnitude bins with $\Delta$m$=$0.05 mag.  The 50\% and 90\% completeness limits were defined as the faintest bin where the completeness exceeded 50\% or 90\%, respectively.  We note that these completeness limits are applicable to regions of average depth in a given tile.  The median completeness limits for the entire survey are: 50\% $=$ 22.65 and 90\% $=$ 22.15.  We first show the completeness curves for all tiles in Figure 14, followed by the 50\% and 90\% completeness limits in Tables 5--8.}

\edit3{Following the completeness tables, we show summary plots and give a summary table for the nine massive quiescent candidates removed following visual inspection.  Finally, In this section we also provide example tables showing the format of the full machine-readable NHS catalog.}

\begin{figure}[!h]
\centering
\includegraphics[scale=0.6,angle=0]{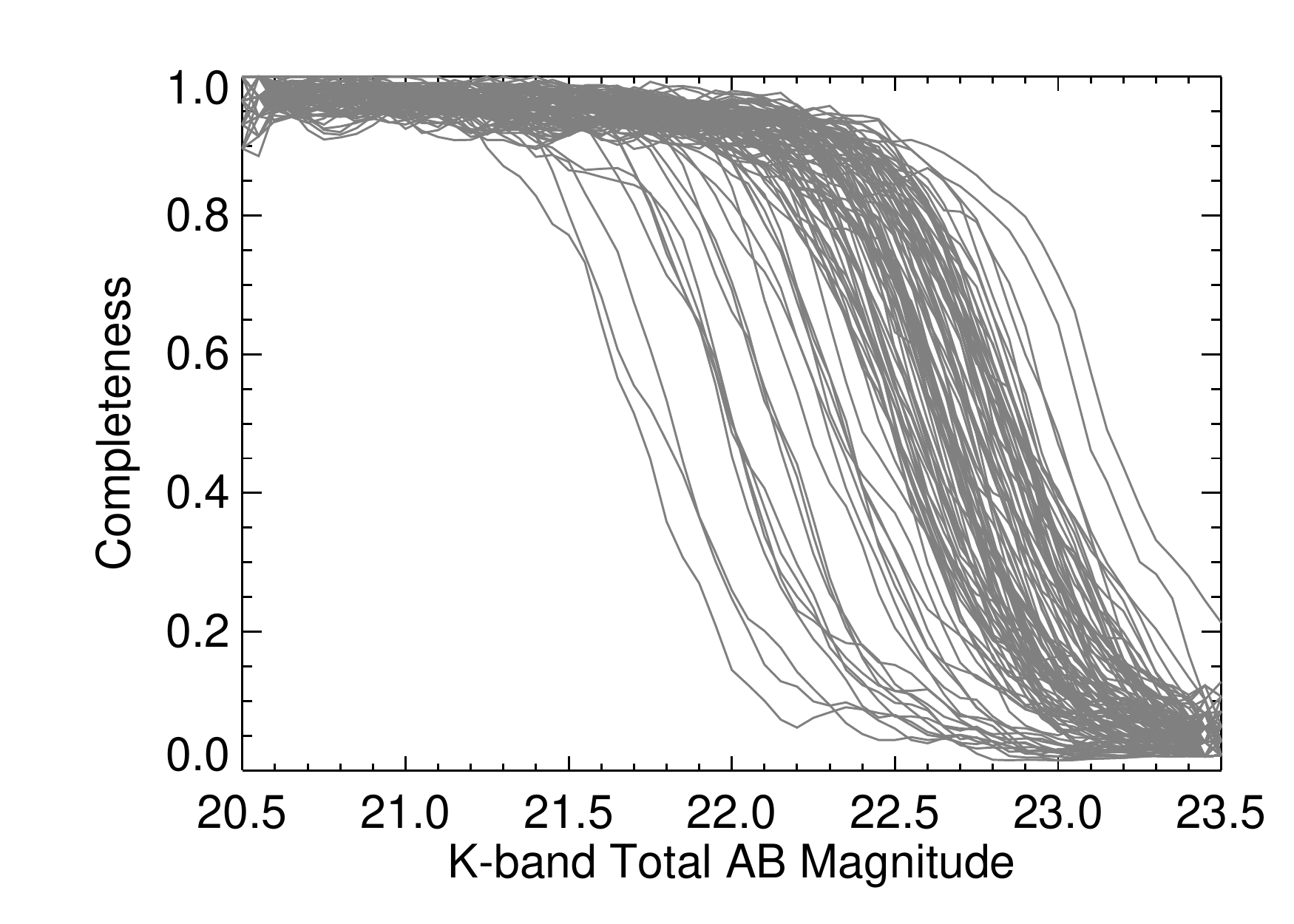}
\caption{Completeness curves derived from mock-source injection simulations into each tile.  The median completeness limits for the entire survey are: 50\% $=$ 22.65 and 90\% $=$ 22.15.}
\label{Fig:allcomp}
\end{figure}

\begin{deluxetable}{ccc}[!b]
\tablecaption{NHS $K$-band Completeness for A Fields} 
\tablehead{\colhead{Field} & \colhead{50\% Completeness} & \colhead{90\% Completeness}}
\startdata 
A1&22.6&22.2\\
A2&22.9&22.3\\
A3&22.8&22.4\\
A4&22.8&22.4\\
A5&22.9&22.4\\
A6&22.9&21.9\\
A7&22.1&21.8\\
A8&22.9&22.4\\
A9&22.6&22.1\\
A10&22.8&22.4\\
A11&22.5&22.1\\
A12&22.5&22.1\\
A13&22.8&22.4\\
A14&22.3&21.8\\
A15&22.7&22.2\\
A16&22.6&22.2\\
A17&22.6&22.3\\
A18&22.9&22.4\\
A19&21.7&21.2\\
A20&22.0&21.6\\
A21&22.2&21.8\\
A22&22.6&22.2\\
A23&21.9&21.6\\
A24&22.9&22.3\\
A25&22.6&22.2\\
A26&22.5&21.9\\
A27&22.4&21.9
\enddata  
\end{deluxetable}

\begin{deluxetable}{ccc} 
\tablecaption{NHS $K$-band Completeness for B Fields} 
\tablehead{\colhead{Field} & \colhead{50\% Completeness} & \colhead{90\% Completeness}}
\startdata 
B1&22.6&22.3\\
B2&22.6&22.1\\
B3&22.8&22.1\\
B4&22.7&22.2\\
B5&22.6&22.1\\
B6&22.9&22.4\\
B7&22.9&22.4\\
B8&22.9&22.4\\
B9&22.8&22.4\\
B10&22.7&22.2\\
B11&22.8&22.4\\
B12&23.1&22.5\\
B13&22.8&22.4\\
B14&22.7&22.2\\
B15&23.1&22.6\\
B16&22.9&22.4\\
B17&22.6&22.3\\
B18&22.4&22.1\\
B19&22.6&22.1\\
B20&22.8&22.2\\
B21&22.8&22.3\\
B22&22.6&22.2\\
B23&22.6&22.2\\
B24&22.6&22.4\\
B25&22.9&22.1\\
B26&22.7&22.4\\
B27&22.9&22.3\\
B28&22.6&22.1
\enddata  
\end{deluxetable}

\begin{deluxetable}{ccc} 
\tablecaption{NHS $K$-band Completeness for C Fields} 
\tablehead{\colhead{Field} & \colhead{50\% Completeness} & \colhead{90\% Completeness}}
\startdata 
C1&22.9&22.5\\
C2&22.5&22.1\\
C3&22.9&22.4\\
C4&22.9&22.4\\
C5&22.6&22.2\\
C6&22.6&22.1\\
C7&22.7&22.1\\
C8&22.8&22.4\\
C9&22.6&22.1\\
C10&22.6&22.1\\
C11&22.6&22.1\\
C12&22.6&22.1\\
C13&22.0&21.4\\
C14&22.8&22.3\\
C15&22.5&22.1\\
C16&22.6&22.2\\
C17&22.5&21.9\\
C18&22.5&22.1\\
C19&22.7&22.4\\
C20&22.6&22.3\\
C21&22.4&21.9\\
C22&22.8&22.4\\
C23&22.5&22.0\\
C24&22.4&22.1\\
C25&22.7&22.4\\
C26&22.6&22.2\\
C27&22.9&22.4\\
C28&22.8&22.3
\enddata  
\end{deluxetable}

\begin{deluxetable}{ccc} 
\tablecaption{NHS $K$-band Completeness for D Fields} 
\tablehead{\colhead{Field} & \colhead{50\% Completeness} & \colhead{90\% Completeness}}
\startdata 
D1&22.4&21.9\\
D2.3&22.6&22.1\\
D2.7&22.4&22.0\\
D2&22.9&22.3\\
D3&22.1&21.8\\
D4.3&22.3&21.9\\
D4.7&22.4&22.0\\
D4&22.4&21.9\\
D5&22.4&22.1\\
D6.3&22.6&22.1\\
D6.7&22.6&22.1\\
D6&22.7&22.1\\
D7&22.3&21.9\\
D8.3&22.3&21.9\\
D8.7&22.2&21.9\\
D8&22.8&22.2\\
D9&22.6&22.1\\
D10&22.6&22.2\\
D10.3&21.8&21.4\\
D10.7&21.9&21.4\\
D11&21.8&21.4\\
D12&22.6&21.9\\
D13&21.9&21.6\\
D14&22.6&22.1\\
D15&22.6&22.2\\
D16&22.1&21.7
\enddata  
\end{deluxetable}

\begin{figure*}[ht]
\centering
\includegraphics[scale=0.6,angle=0]{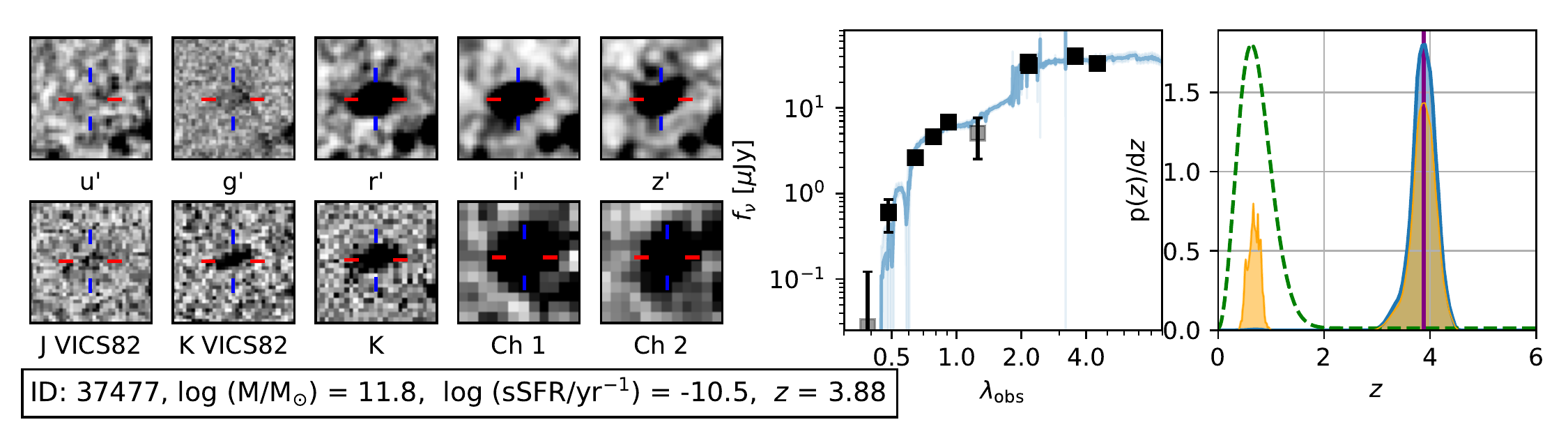}
\includegraphics[scale=0.6,angle=0]{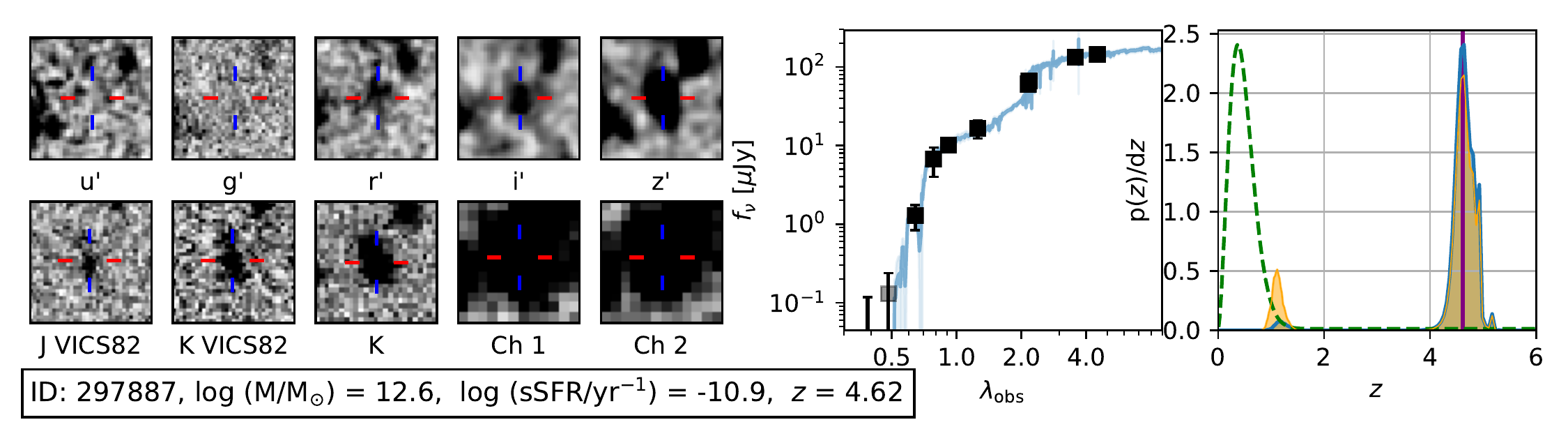}
\includegraphics[scale=0.6,angle=0]{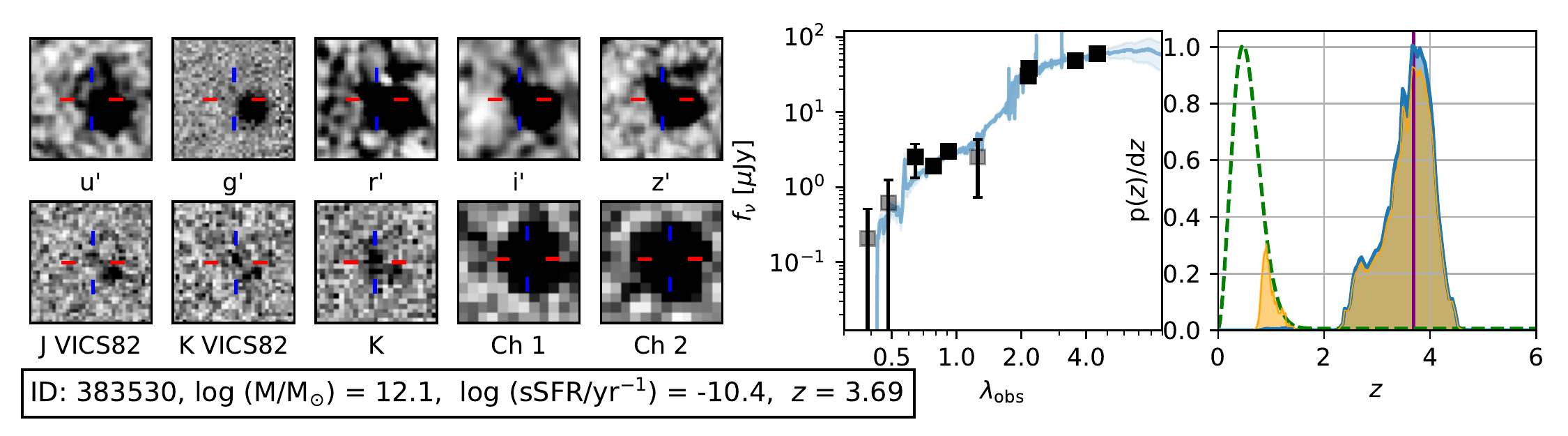}
\includegraphics[scale=0.6,angle=0]{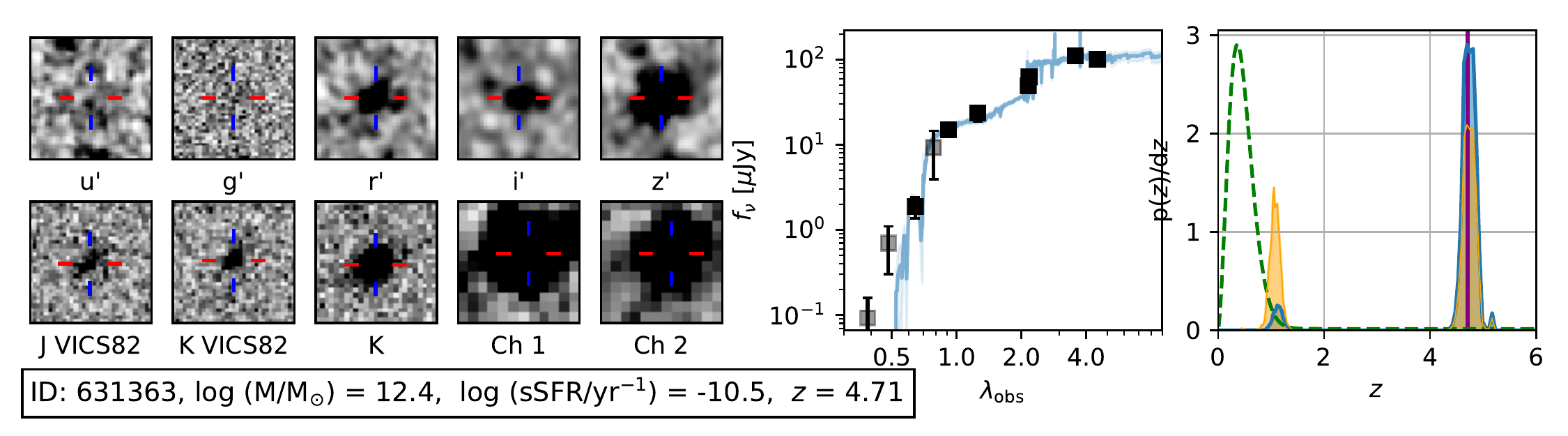}
\includegraphics[scale=0.6,angle=0]{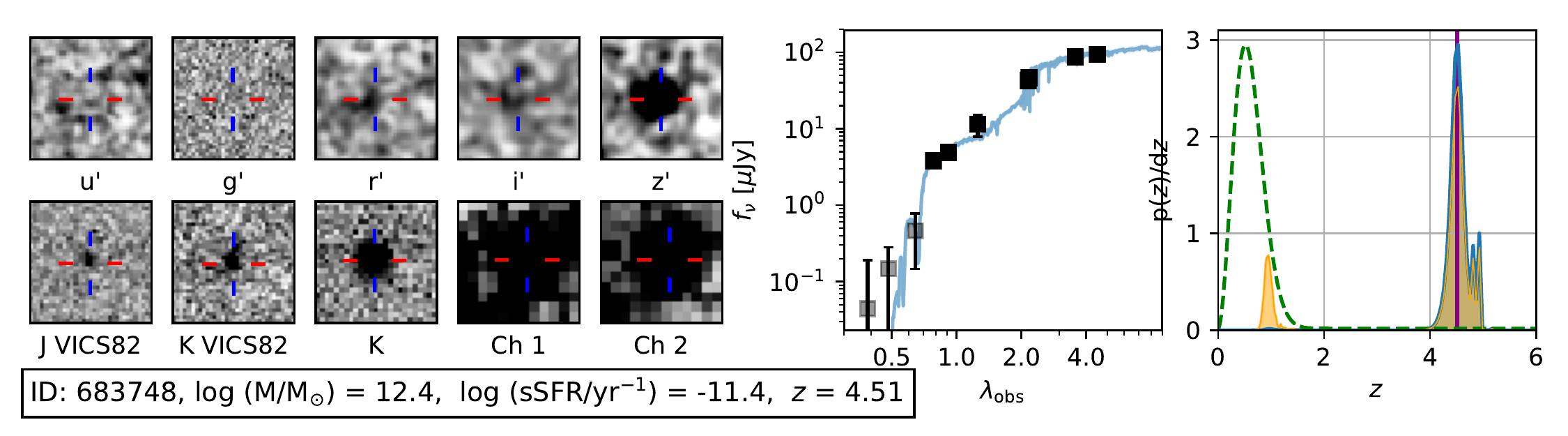}
\caption{Summary plots of the first five of nine candidate massive quiescent galaxies removed following visual inspection.  All symbols are the same as Figure~\ref{Fig:stamps-q}}.
\label{Fig:stamps-q-appen1}
\end{figure*}

\begin{figure*}[!h]
\centering
\includegraphics[scale=0.64,angle=0]{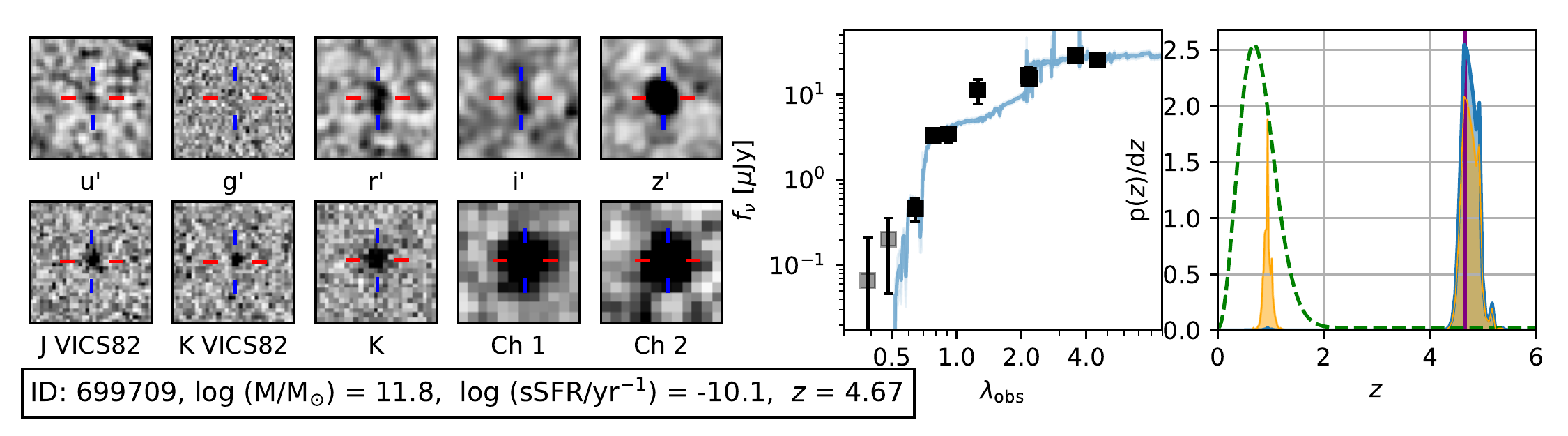}
\includegraphics[scale=0.64,angle=0]{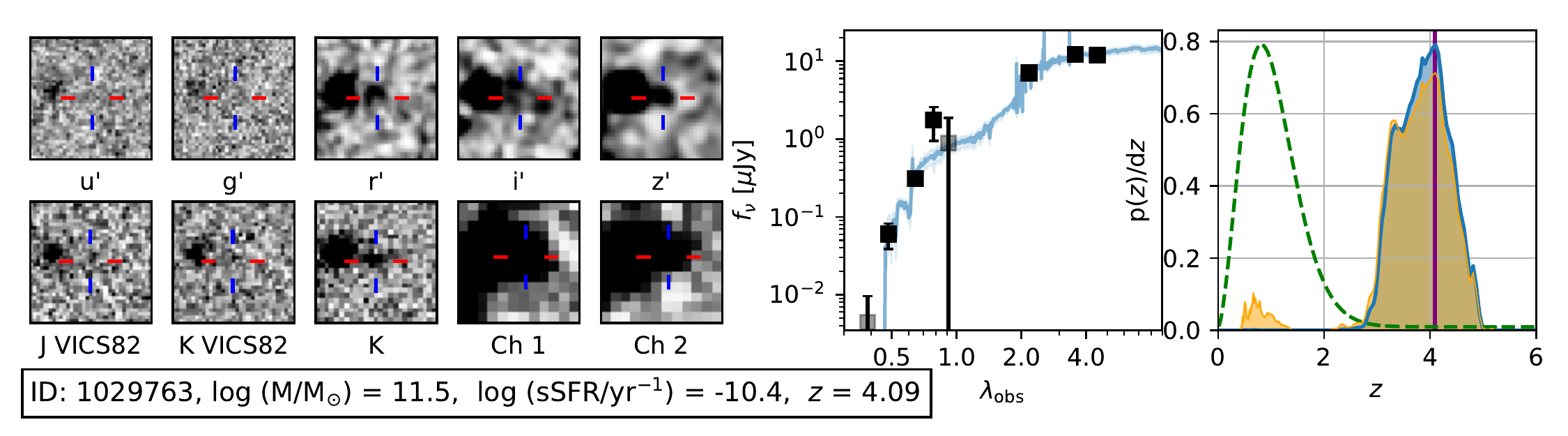}
\includegraphics[scale=0.64,angle=0]{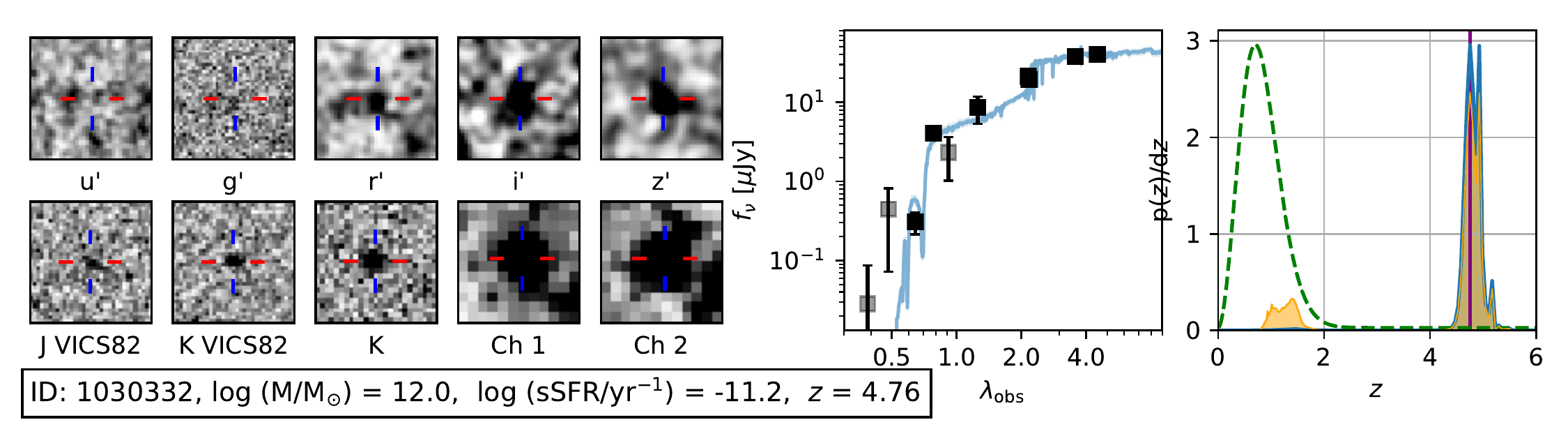}
\includegraphics[scale=0.64,angle=0]{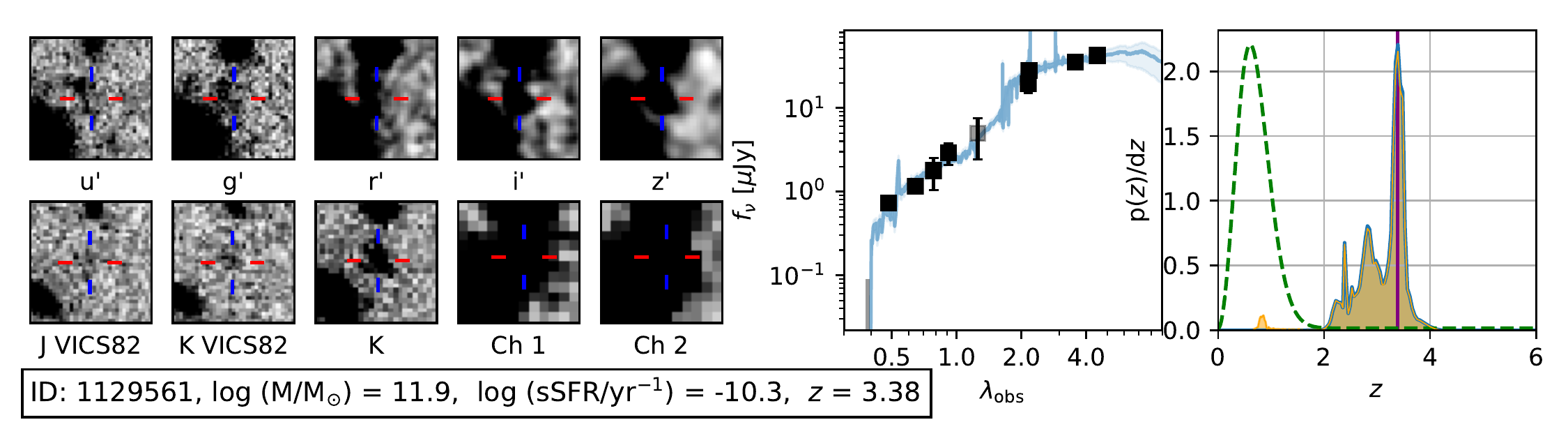}
\caption{Summary plots of the last four of nine candidate massive quiescent galaxies removed following visual inspection.  All symbols are the same as Figure~\ref{Fig:stamps-q}}.
\label{Fig:stamps-q-appen2}
\end{figure*}

\begin{deluxetable*}{cccccccccccc}[!h]
\tablecaption{Candidate Massive Quiescent Galaxies Removed Following Visual Inspection} 
\tablehead{\colhead{ID} & \colhead{RA} & \colhead{Dec} & \colhead{$K_{AB}$} & \colhead{$z_{p}$} & \colhead{$\int P(z_{p}) >$ 3} & \colhead{log(M$_{\ast}$)} & \colhead{log(sSFR)} & \colhead{log(M$_{\ast}$)} & \colhead{log sSFR} & \colhead{$\int$P(sSFR)$<-$10.05}\\
\colhead{} & \colhead{(J2000)} & \colhead{(J2000)} & \colhead{} & \colhead{} & \colhead{} & \colhead{EAZY} & \colhead{EAZY} & \colhead{BC03} & \colhead{BC03} & \colhead{BC03}\\
\colhead{} & \colhead{(deg)} & \colhead{(deg)} & \colhead{(mag)} & \colhead{$ $} & \colhead{$ $} & \colhead{(M\sol)} & \colhead{(yr$^{-1}$)} & \colhead{(M\sol)} & \colhead{(yr$^{-1}$)} & \colhead{$ $}}
\startdata 
37477$^{s}$&15.122565&0.086271&20.2&3.88&0.79&11.8&-10.5&11.7&$<$-11.0&0.79\\
297887$^{s}$&16.585555&0.138326&19.3&4.62&0.87&12.6&-10.9&12.5&$<$-11.8&0.96\\
383530$^{s}$&17.528702&0.607760&19.9&3.69&0.80&12.1&-10.4&12.1&-10.4&0.87\\
631363$^{u}$&19.715019&0.215609&19.4&4.71&0.68&12.4&-10.5&12.3&$<$-11.6&0.85\\
683748$^{s}$&19.803261&0.133706&19.7&4.51&0.85&12.4&-11.4&12.4&$<$-11.7&0.98\\
699709$^{u}$&19.762647&-0.210499&20.9&4.67&0.82&11.8&-10.1&11.8&$<$-11.1&0.76\\
1029763$^{v}$&22.549563&0.528559&21.8&4.09&0.93&11.5&-10.4&11.5&-10.1&0.89\\
1030332$^{u}$&22.401811&0.542126&20.7&4.76&0.83&12.0&-11.2&11.9&$<$-11.2&0.85\\
1129561$^{u}$&23.194968&-0.730294&20.3&3.38&0.63&11.9&-10.3&11.8&-10.1&0.82\\
\enddata  
\tablecomments{Properties of the 9 candidate massive quiescent galaxies removed following visual inspection.  $^{u}$Removed due to visible weak flux in the $u$-band image coincident with the position of the source.  $^{s}$ Removed due to visbly appearing elongated with measured r$_h$ indicating the source being significantly resolved in the $K_s$-band imaging ($>$1$^{\prime\prime}$.  $^{v}$Removed due to lack of measured VISTA $J$ and $K$ photometry.}
\label{qgcat-appen}
\end{deluxetable*}

\begin{rotatetable*}
\begin{deluxetable*}{ccccccccccc} 
\tabletypesize{\scriptsize} \setlength{\tabcolsep}{0.02in}
\tablecolumns{11} 
\tablewidth{0pc} 
\tablecaption{NHS Catalog Sample} 
\tablehead{ 
\colhead{ID} & \colhead{R.A.(J2000)} & \colhead{decl.(J2000)} & \colhead{$a$} & \colhead{$e$} & \colhead{$\theta$} & \colhead{$r_{1/2}$} & \colhead{$f_{K_{s},\textrm{NHS}}^{\textrm{AUTO}}$} & \colhead{$\sigma_{K_{s},\textrm{NHS}}^{\textrm{AUTO}}$} & \colhead{$f_{K_{s},\textrm{NHS}}^{\textrm{OPT}}$} & \colhead{$\sigma_{K_{s},\textrm{NHS}}^{\textrm{OPT}}$} \\
\colhead{} & \colhead{(deg)} & \colhead{(deg)} & \colhead{(arcsec)} & \colhead{} & \colhead{(deg)} & \colhead{(arcsec)} & \colhead{($\mu$Jy)} & \colhead{($\mu$Jy)} & \colhead{($\mu$Jy)} & \colhead{($\mu$Jy)} \\ 
\colhead{(1)} & \colhead{(2)} & \colhead{(3)} & \colhead{(4)} & \colhead{(5)} & \colhead{(6)} & \colhead{(7)} & \colhead{(8)} & \colhead{(9)} & \colhead{(10)} & \colhead{(11)}  
}
\startdata 
1100000 & 23.192657 & 0.257515 & 0.0 & 0.15 & 34.1 & 3.3 & 460.491 & 1.996 & 252.966 & 0.718 \\
1100001 & 22.954510 & 0.260587 & 0.0 & 0.33 & 66.7 & 2.9 & 18.876 & 1.079 & 12.031 & 0.747 \\
1100002 & 22.914940 & 0.261128 & 0.0 & 0.08 & -1.6 & 2.1 & 58.961 & 0.983 & 62.439 & 0.730 \\
1100003 & 23.125761 & 0.262926 & 0.0 & 0.34 & 9.0 & 3.0 & 5.419 & 0.456 & 4.052 & 0.694 \\
1100004 & 23.200405 & 0.262837 & 0.0 & 0.43 & 58.5 & 1.9 & 2.616 & 0.282 & 2.964 & 0.711 \\
1100005 & 22.951833 & 0.262902 & 0.0 & 0.24 & 53.5 & 1.9 & 4.060 & 0.523 & 4.840 & 0.730 \\
1100006 & 23.151630 & 0.262835 & 0.0 & 0.04 & 56.1 & 1.9 & 3.277 & 0.424 & 3.796 & 0.694 \\
1100007 & 22.962145 & 0.262063 & 0.0 & 0.24 & 1.8 & 2.4 & 19.283 & 0.928 & 17.268 & 0.745 \\
1100008 & 23.173936 & 0.262909 & 0.0 & 0.31 & -22.2 & 1.8 & 4.517 & 0.554 & 5.727 & 0.708 \\
1100009 & 23.151618 & 0.260111 & 0.0 & 0.13 & -35.5 & 2.8 & 84.666 & 1.338 & 58.754 & 0.696 \\
\enddata  
\tablecomments{Table 4 is published in its entirety in the machine-readable format. A portion is shown here for guidance regarding its form and content. (1) Unique object ID number, (2) object R.A. (J2000) in decimal degrees, (3) object decl. (J2000) in decimal degrees, (4) semimajor axis in the detection image, (5) ellipticity measured in the detection image, defined as $e=1-b/a$, where $b$  and $a$ are the semiminor and semimajor axes, respectively, (6) position angle measured in the detection image, (7) \pkg{Source Extractor} half-light radius (FLUX\_RADIUS),
(8) NHS $K_{s}$ Kron fluxes corrected to total, (9) uncertainties in NHS $K_{s}$ Kron fluxes corrected to total, (10) NHS $K_{s}$ optimal-aperture fluxes corrected to total, (11) NHS $K_{s}$ optimal-aperture fluxes corrected to total, (12,13,14,15) NHS field labels containing source, (16) maximum pixel value in the isophotal area of the source in the data quality mask image, (17,22,27,32,37) DECam $ugriz$ \pkg{Tractor} \edit2{total} fluxes, (18,23,28,33,38) ucertainties in DECam $ugriz$ \pkg{Tractor} \edit2{total} fluxes, (19,24,29,34,39) DECam $ugriz$ \pkg{Tractor} model profiles, (20,25,30,25,40) DECam $ugriz$ \pkg{Tractor} log-probabilities, (21,26,31,36,41) DECam $ugriz$ \pkg{Tractor} flags, (42) $JK$ object ID number from \citet{geach17}, (43,45) $JK$ Kron total fluxes from \citet{geach17}, (44,46) derived uncertainties in $JK$ Kron total fluxes, (47) Flag for VICS82 sources with values of zero in isophotal area of the source in the WTM, (48,53) 3.6 and 4.5 $\mu$m IRAC \pkg{Tractor} total fluxes, (49,54)  uncertainties in 3.6 and 4.5 $\mu$m IRAC \pkg{Tractor} total fluxes, (50,55)  3.6 and 4.5 $\mu$m IRAC \pkg{Tractor} model profiles, (51,56) 3.6 and 4.5 $\mu$m IRAC \pkg{Tractor} log-probabilities, (52,57) 3.6 and 4.5 $\mu$m IRAC \pkg{Tractor} flags, (58) SDSS spectroscopic redshift, (59) \pkg{eazy-py} photometric redshift, (60)  \pkg{eazy-py} photometric redshift $\chi^{2}$ minimum value, (61) Number of peaks detected in P(z) from \pkg{eazy-py}, (62) number of filters used in \pkg{eazy-py} photometric redshift estimate, (63) log stellar mass from \pkg{eazy-py}, (64) log SFR from \pkg{eazy-py}, (65,66,67) Rest-frame $UVJ$  magnitudes from \pkg{eazy-py}.}
\label{mcat1}
\end{deluxetable*}
\end{rotatetable*}

\begin{rotatetable*}
\addtocounter{table}{-1}
\begin{deluxetable*}{ccccccccccc} 
\tabletypesize{\scriptsize} \setlength{\tabcolsep}{0.02in}
\tablecolumns{11} 
\tablewidth{0pc} 
\tablecaption{NHS Catalog Sample (continued)} 
\tablehead{ 
\colhead{NHS Field A} & \colhead{NHS Field B} & \colhead{NHS Field C} & \colhead{NHS Field D} & \colhead{External $K_{s},\textrm{NHS}$ Flag} & \colhead{$f_{u}$} & \colhead{$\sigma_{u}$} & \colhead{$\textrm{model}_u$} & \colhead{$\textrm{log(}P_{u}\textrm{)}$} & \colhead{Tractor $u'$ Flag} & \colhead{$f_{g}$} \\
\colhead{} & \colhead{} & \colhead{} & \colhead{} & \colhead{} & \colhead{($\mu$Jy)} & \colhead{($\mu$Jy)} & \colhead{} & \colhead{} & \colhead{} & \colhead{($\mu$Jy)} \\ 
\colhead{(12)} & \colhead{(13)} & \colhead{(14)} & \colhead{(15)} & \colhead{(16)} & \colhead{(17)} & \colhead{(18)} & \colhead{(19)} & \colhead{(20)} & \colhead{(21)} & \colhead{(22)}  
}
\startdata 
B20 & N/A & N/A & N/A & 0 & 6.090 & 0.454 & 4 & -794705.50 & 0 & 26.644 \\
B20 & N/A & N/A & N/A & 0 & 1.626 & 0.196 & 1 & -16582126.00 & 0 & 3.420 \\
B20 & N/A & N/A & N/A & 0 & 0.374 & 0.076 & 1 & -751686.50 & 0 & 2.279 \\
B20 & N/A & N/A & N/A & 0 & 2.369 & 0.200 & 4 & -686279.50 & 0 & 1.867 \\
B20 & N/A & N/A & N/A & 0 & 0.516 & 0.193 & 4 & -773428.31 & 0 & 0.792 \\
B20 & N/A & N/A & N/A & 0 & 0.523 & 0.154 & 1 & -982226.38 & 0 & 0.773 \\
B20 & N/A & N/A & N/A & 0 & 0.667 & 0.054 & 0 & -995190.56 & 0 & 0.817 \\
B20 & N/A & N/A & N/A & 0 & 0.024 & 0.080 & 0 & -719684.38 & 0 & 0.122 \\
B20 & N/A & N/A & N/A & 0 & 0.946 & 0.219 & 4 & -656113.25 & 0 & 0.963 \\
B20 & N/A & N/A & N/A & 0 & -0.075 & 0.090 & 1 & -836314.19 & 0 & 1.022 \\
\enddata  
\label{mcat2}
\end{deluxetable*}
\end{rotatetable*}

\begin{rotatetable*}
\addtocounter{table}{-1}
\begin{deluxetable*}{ccccccccccc} 
\tabletypesize{\scriptsize} \setlength{\tabcolsep}{0.02in}
\tablecolumns{11} 
\tablewidth{0pc} 
\tablecaption{NHS Catalog Sample (continued)} 
\tablehead{ 
\colhead{$\sigma_{g}$} & \colhead{$\textrm{model}_g$} & \colhead{$\textrm{log(}P_{g}\textrm{)}$} & \colhead{Tractor $g'$ Flag} & \colhead{$f_{r}$} & \colhead{$\sigma_{r}$} & \colhead{$\textrm{model}_r$} & \colhead{$\textrm{log(}P_{r}\textrm{)}$} & \colhead{Tractor $r'$ Flag} & \colhead{$f_{i}$} & \colhead{$\sigma_{i}$} \\
\colhead{($\mu$Jy)} & \colhead{} & \colhead{} & \colhead{} & \colhead{($\mu$Jy)} & \colhead{($\mu$Jy)} & \colhead{} & \colhead{} & \colhead{} & \colhead{($\mu$Jy)} & \colhead{($\mu$Jy)} \\ 
\colhead{(23)} & \colhead{(24)} & \colhead{(25)} & \colhead{(26)} & \colhead{(27)} & \colhead{(28)} & \colhead{(29)} & \colhead{(30)} & \colhead{(31)} & \colhead{(32)} & \colhead{(33)}  
}
\startdata 
1.000 & 4 & -1498032.38 & 0 & 84.053 & 2.769 & 1 & -609655.19 & 0 & 135.279 & 4.470 \\
0.311 & 1 & -36960160.00 & 0 & 7.465 & 0.739 & 1 & -3387295.75 & 0 & 9.377 & 1.168 \\
0.102 & 0 & -1453222.62 & 0 & 9.256 & 0.706 & 4 & -179494.30 & 0 & 30.529 & 1.278 \\
0.090 & 0 & -1360433.12 & 0 & 3.711 & 0.956 & 4 & -205406.78 & 0 & 2.718 & 0.821 \\
0.207 & 1 & -1342367.88 & 0 & 1.008 & 0.523 & 1 & -173332.89 & 0 & 0.452 & 0.841 \\
0.442 & 4 & -1842914.88 & 0 & 1.165 & 0.609 & 1 & -441678.34 & 0 & 0.265 & 0.411 \\
0.207 & 4 & -3033535.00 & 0 & 0.830 & 0.356 & 1 & -6063350.00 & 0 & 0.992 & 0.319 \\
0.206 & 0 & -1384284.88 & 0 & 0.277 & 0.706 & 4 & -139799.28 & 0 & 1.478 & 0.978 \\
0.279 & 4 & -1345640.38 & 0 & 1.464 & 0.466 & 1 & -183884.48 & 0 & 1.916 & 0.944 \\
0.122 & 4 & -1535087.88 & 0 & 5.400 & 0.292 & 1 & -747731.94 & 0 & 21.060 & 2.221 \\
\enddata  
\label{mcat3}
\end{deluxetable*}
\end{rotatetable*}

\clearpage 
\begin{rotatetable*}
\addtocounter{table}{-1}
\begin{deluxetable*}{ccccccccccc} 
\tabletypesize{\scriptsize} \setlength{\tabcolsep}{0.02in}
\tablecolumns{11} 
\tablewidth{0pc} 
\tablecaption{NHS Catalog Sample (continued)} 
\tablehead{ 
\colhead{$\textrm{model}_i$} & \colhead{$\textrm{log(}P_{i}\textrm{)}$} & \colhead{Tractor $i'$ Flag} & \colhead{$f_{z}$} & \colhead{$\sigma_{z}$} & \colhead{$\textrm{model}_z$} & \colhead{$\textrm{log(}P_{z}\textrm{)}$} & \colhead{Tractor $z'$ Flag} & \colhead{VICS82ID} & \colhead{$f_{{J},\textrm{VICS82}}^{\textrm{AUTO}}$} & \colhead{$\sigma_{{J},\textrm{VICS82}}^{\textrm{AUTO}}$} \\
\colhead{} & \colhead{} & \colhead{} & \colhead{($\mu$Jy)} & \colhead{($\mu$Jy)} & \colhead{} & \colhead{} & \colhead{} & \colhead{(Geach+17)} & \colhead{($\mu$Jy)} & \colhead{($\mu$Jy)} \\ 
\colhead{(34)} & \colhead{(35)} & \colhead{(36)} & \colhead{(37)} & \colhead{(38)} & \colhead{(39)} & \colhead{(40)} & \colhead{(41)} & \colhead{(42)} & \colhead{(43)} & \colhead{(44)}  
}
\startdata 
1 & -419193.06 & 0 & 179.684 & 5.904 & 1 & -318709.12 & 0 & VICS82J013246.24+001527.0 & 284.44 & 10.88 \\
1 & -822670.38 & 0 & 12.094 & 1.177 & 1 & -192591.42 & 0 & VICS82J013149.10+001538.6 & 13.89 & 3.89 \\
1 & -188672.66 & 0 & 49.241 & 1.789 & 1 & -128849.97 & 0 & VICS82J013139.59+001540.3 & 66.51 & 6.16 \\
1 & -166728.31 & 0 & 3.377 & 0.822 & 1 & -122933.82 & 0 & -99 & -99.00 & -99.00 \\
1 & -183598.70 & 0 & 2.044 & 1.239 & 1 & -104748.86 & 0 & -99 & -99.00 & -99.00 \\
1 & -344773.72 & 0 & 1.909 & 1.167 & 1 & -411497.25 & 0 & VICS82J013148.47+001546.7 & -99.00 & -99.00 \\
1 & -4665033.00 & 0 & 1.281 & 0.781 & 1 & -10052148.00 & 0 & VICS82J013236.42+001546.5 & -99.00 & -99.00 \\
1 & -163797.36 & 0 & 3.162 & 1.424 & 1 & -97862.48 & 0 & VICS82J013150.91+001543.5 & 5.06 & 2.52 \\
1 & -191571.53 & 0 & 2.853 & 0.872 & 1 & -187026.06 & 0 & -99 & -99.00 & -99.00 \\
4 & -423303.53 & 0 & 23.014 & 1.260 & 1 & -638833.19 & 0 & VICS82J013236.40+001536.4 & 31.67 & 5.11 \\
\enddata  
\label{mcat4}
\end{deluxetable*}
\end{rotatetable*}
\clearpage

\clearpage 
\begin{rotatetable*}
\addtocounter{table}{-1}
\begin{deluxetable*}{ccccccccccc} 
\tabletypesize{\scriptsize} \setlength{\tabcolsep}{0.02in}
\tablecolumns{11} 
\tablewidth{0pc} 
\tablecaption{NHS Catalog Sample (continued)} 
\tablehead{ 
\colhead{$f_{K_{s},\textrm{VICS82}}^{\textrm{AUTO}}$} & 
\colhead{$\sigma_{K_{s},\textrm{VICS82}}^{\textrm{AUTO}}$} & 
\colhead{VICS82 WTM Flag} & 
\colhead{$f_{\textrm{3.6}}$} & 
\colhead{$\sigma_{\textrm{3.6}}$} & 
\colhead{$\textrm{model}_{\textrm{3.6}}$} & 
\colhead{$\textrm{log(}P_{\textrm{3.6}}\textrm{)}$} & 
\colhead{Tractor 3.6 Flag} & 
\colhead{$f_{\textrm{4.5}}$} & 
\colhead{$\sigma_{\textrm{4.5}}$} & 
\colhead{$\textrm{model}_{\textrm{4.5}}$}
 \\
\colhead{($\mu$Jy)} & \colhead{($\mu$Jy)} & \colhead{} & \colhead{($\mu$Jy)} & \colhead{($\mu$Jy)} & \colhead{} & \colhead{} & \colhead{} & \colhead{($\mu$Jy)} & \colhead{($\mu$Jy)} & \colhead{} \\ 
\colhead{(45)} & \colhead{(46)} & \colhead{(47)} & \colhead{(48)} & \colhead{(49)} & \colhead{(50)} & \colhead{(51)} & \colhead{(52)} & \colhead{(53)} & \colhead{(54)} & \colhead{(55)}  
}
\startdata 
463.37 & 16.42 & 0 & 257.56 & 12.76 & 4 & -8.01 & 0 & 216.08 & 10.81 & 4 \\
10.07 & 3.43 & 0 & 7.82 & 1.47 & 0 & -2.80 & 0 & 8.07 & 1.57 & 0 \\
64.51 & 6.12 & 0 & 36.08 & 2.24 & 0 & -2.51 & 0 & 24.17 & 1.91 & 0 \\
-99.00 & -99.00 & 0 & 4.70 & 1.17 & 0 & -2.15 & 0 & 3.50 & 0.82 & 1 \\
-99.00 & -99.00 & 0 & 5.77 & 1.75 & 0 & -2.05 & 0 & 4.60 & 1.62 & 0 \\
9.19 & 3.30 & 0 & 8.79 & 0.85 & 0 & -3.10 & 0 & 9.18 & 0.89 & 0 \\
3.02 & 1.95 & 0 & 2.95 & 1.32 & 0 & -19.37 & 0 & 2.64 & 1.45 & 0 \\
23.14 & 4.65 & 0 & 29.31 & 2.00 & 0 & -2.72 & 0 & 31.09 & 2.03 & 0 \\
-99.00 & -99.00 & 0 & 5.07 & 1.67 & 0 & -5.73 & 0 & 3.00 & 1.70 & 0 \\
68.62 & 6.21 & 0 & 96.07 & 5.08 & 1 & -17.76 & 0 & 51.32 & 3.01 & 0 \\
\enddata  
\label{mcat5}
\end{deluxetable*}
\end{rotatetable*}
\clearpage

\clearpage 
\begin{rotatetable*}
\addtocounter{table}{-1}
\begin{deluxetable*}{cccccccccccc} 
\tabletypesize{\scriptsize} \setlength{\tabcolsep}{0.02in}
\tablecolumns{12} 
\tablewidth{0pc} 
\tablecaption{NHS Catalog Sample (continued)} 
\tablehead{ 
\colhead{$\textrm{log(}P_{\textrm{4.5}}\textrm{)}$} & \colhead{Tractor 4.5 Flag} & \colhead{$z_{\textrm{spec}}$} & \colhead{$z_{\textrm{phot}}$} & \colhead{$\chi^{2}_{\textrm{phot}}$} & \colhead{$N_{\textrm{peaks}}$} & \colhead{$N_{\textrm{filters}}$} & \colhead{log(M$_{*}$)} & \colhead{log(SFR)} & \colhead{\edit2{$f_{\textrm{U,rest}}$}} & \colhead{\edit2{$f_{\textrm{V,rest}}$}} & \colhead{\edit2{$f_{\textrm{J,rest}}$}} \\
\colhead{} & \colhead{} & \colhead{(SDSS)} & \colhead{(\pkg{eazy-py})} & \colhead{(\pkg{eazy-py})} & \colhead{(\pkg{eazy-py})} & \colhead{(\pkg{eazy-py})} & \colhead{(M$_{\odot}$)} & \colhead{(M$_{\odot}$ yr$^{-1}$)} & \colhead{\edit2{($\mu$Jy)}} & \colhead{\edit2{($\mu$Jy)}} & \colhead{\edit2{($\mu$Jy)}} \\ 
\colhead{(56)} & \colhead{(57)} & \colhead{(58)} & \colhead{(59)} & \colhead{(60)} & \colhead{(61)} & \colhead{(62)} & \colhead{(63)} & \colhead{(64)} & \colhead{(65)} & \colhead{(66)} & \colhead{(67)}  
}
\startdata 
-6.22 & 0 & -99.0000 & 0.33 & 1.26 & 1 & 10 & 11.1 & 0.2 & 22.3 & 122.5 & 378.8 \\
-2.23 & 0 & -99.0000 & 0.30 & 4.59 & 1 & 10 & 9.3 & -0.4 & 2.9 & 8.3 & 15.6 \\
-2.47 & 0 & -99.0000 & 0.39 & 43.61 & 1 & 10 & 10.6 & -2.0 & 2.7 & 21.6 & 60.3 \\
-2.14 & 0 & -99.0000 & 0.64 & 2.78 & 2 & 8 & 9.1 & 0.5 & 2.5 & 3.5 & 5.3 \\
-1.88 & 0 & -99.0000 & 2.58 & 1.46 & 2 & 8 & 10.2 & 1.7 & 1.5 & 2.4 & 5.1 \\
-2.54 & 0 & -99.0000 & 2.46 & 3.26 & 1 & 9 & 10.5 & 1.8 & 1.4 & 3.3 & 9.2 \\
-10.77 & 0 & -99.0000 & 2.09 & 0.29 & 6 & 9 & 10.2 & 0.8 & 1.6 & 2.9 & 3.1 \\
-2.22 & 0 & -99.0000 & 1.30 & 1.68 & 2 & 10 & 10.8 & 1.8 & 1.3 & 6.5 & 25.8 \\
-3.45 & 0 & -99.0000 & 0.74 & 0.05 & 3 & 8 & 9.3 & 0.3 & 1.4 & 2.1 & 4.5 \\
-10.75 & 0 & -99.0000 & 0.63 & 19.84 & 1 & 10 & 11.1 & -1.6 & 3.1 & 21.7 & 75.3 \\
\enddata  
\label{mcat6}
\end{deluxetable*}
\end{rotatetable*}
\clearpage

\end{document}